\documentclass[11pt]{article}
\usepackage[utf8]{inputenc}
\usepackage[T1]{fontenc}
\usepackage{lmodern}
\usepackage[margin=1in]{geometry}
\usepackage{microtype}
\usepackage{amsmath,amssymb,amsthm}
\usepackage{graphicx}
\usepackage{booktabs}
\usepackage{multirow}
\usepackage{subcaption}
\usepackage{natbib}
\usepackage[title,toc,titletoc,page]{appendix}
\usepackage{authblk}
\usepackage{placeins}
\usepackage[colorlinks=true,linkcolor=blue,citecolor=blue,urlcolor=blue]{hyperref}
\let\OldSubsection\subsection
\renewcommand{\subsection}{\FloatBarrier\OldSubsection}
\newenvironment{acknowledgement}{\section*{Acknowledgements}}{}

\title{Who blocks whom? Probabilistic pass-blocking assignments for evaluating blockers and pass rushers in American football}
\author[1]{Abhijit Brahme\thanks{Corresponding author.}}
\author[2]{Ishan Mehta}
\author[3]{Gregory J. Matthews}
\author[4]{Alexander Franks}
\affil[1]{University of California Santa Barbara, Probability and Statistics, e-mail: abhijitbrahme@ucsb.edu}
\affil[2]{e-mail: ishanumehta13@gmail.com}
\affil[3]{Loyola University Chicago,
 Department of Mathematics and Statistics,
 e-mail: gmatthews1@luc.edu}
\affil[4]{University of California Santa Barbara, Probability and Statistics, e-mail: amfranks@ucsb.edu}
\date{}

\begin{document}
\maketitle

\begin{abstract}
Historically, statistical analysis of offensive lineman has been hindered by the lack of easily measurable quantities.  More recently, with the introduction of player tracking data new methodological advances are now possible.  Using high-dimensional spatio-temporal data, we adapt the defensive-matchup hidden Markov model of \cite{franks2015characterizing} from basketball to football pass protection, producing frame-by-frame probabilistic assignments of each pass blocker to the rushers. We show how this probabilistic assignment is a usable modeling artifact that augments existing player-evaluation frameworks. We directly quantify the attention a rusher commands, upgrade adjusted plus-minus \citep{Macdonald+2012} from all-or-nothing stints to partial, continuous blocking credit in continuous time, yield block-shedding survival metrics, and measure the space a rusher generates for his teammates. Fit to the first eight weeks of the 2021 NFL season, the resulting metrics recover widely-recognized elite rushers and pass protectors and align with independent charting.
\end{abstract}


\section{Introduction} \label{introduction}
American football is a complex game that consists of many different positions, each given a high specialized role:  while there are only eleven athletes per team on the field at a time, there are over two dozen distinct positions. As such, measuring individual performance requires different metrics for each position, each of which targets the tasks assigned to that role.  Historically, statistics were mainly kept for players in the so-called ``skill positions" (i.e., quarterback, running backs, wide receivers, with tight ends often also included in this group). Running backs and wide receivers are evaluated based on yards gained, and quarterbacks are evaluated based on statistics such as completion percentage, yards per attempt, and quarterback rating.  Defensive players are judged on statistics such as sacks, pressures, tackles, and interceptions, but these are relatively recent additions to the collected statistics, with sacks not being officially recorded by the NFL until 1982 \citep{treacy2026sack}, for example.  Kickers and punters are measured by their field goal percentage and average hang time, respectively.  One can see that with the extreme specialization that is inherent to American football, the statistics to measure the performance of each of these positions follow suit.  

While many of these statistics have been collected since the inception of the game, for much of the game's history there have been positions where collection of statistical performance measures have been scant or even non existent. Defensive linemen, and even more so for offensive linemen, have almost no historical statistical data collected for their performance as their contribution to any given play is difficult to quantify in a simple manner.  With the advent of advance tracking data, however, we propose a novel method for analyzing the performance of both offensive and defensive linemen performance.  

On passing plays, the role of the defensive line (and other players who are ``rushers'' on a particular play) is ultimately to get to the quarterback as quickly as possible with the ideal outcome from a defensive rusher's perspective being a sack (i.e., tackling the quarterback for a loss before they are able to throw the ball). 

On the other side of the ball, the goal of the offensive lineman is to block the defensive rushers to allow their quarterback time to throw the ball unimpeded. However, offensive linemen, one of the most important and highest paid position groups \citep{atienza2026paid}, lack reliable measures of their performance. Commonly, the offensive linemen are typically noticed when they err (i.e., they allow a sack or miss a block), events which are relatively rare. Metrics based on counts of these rare events tend to be unreliable estimates of ability. Metrics, such as the one we present here, which better quantify the performance of offensive linemen would be invaluable for coaches and fans alike.

While there are some metrics for quantifying the performance of offensive linemen, most of these measures, focus on the offensive line as a unit rather than individual performance.  For instance, Football Outsiders measures the ability of pass blockers with adjusted sack rate \citep{fo2023glossary}. 
\cite{Burke2018Created} proposes pass block win rate (PBWR), which measures the proportion of time that a pass blocker is able to hold their block for at least 2.5 seconds.  Unlike adjusted sack rate, this measure evaluates individual linemen rather than the entire offensive line as a unit. However, this method may not account for the ability of the rusher, or changing blocking assignments over the course of the play. For example, a blocker may stunt, or pass off a rusher to a teammate, disrupting the metric by incorrectly distributing credit among the blockers responsible for the block.  

Existing work analyzing the performance of offensive linemen involves hand-collecting data on whether a blocker held their block long enough for the QB to throw the football. \citet{AlamarGould2008} used a three-game sample of data from the 2007 season to establish a relationship between the amount of time a lineman can hold a block and pass completion. Later, \cite{AlamarGoldner2011} looked at passing plays in the 2010 season and estimated the number of yards each lineman contributed on offense. Ultimately, \cite{AlamarGould2008} noted the ``absence of publicly available data that is needed in order to track the performance of a specific lineman'', leading to the lack of research on this topic.  Fortunately, with publicly-available player tracking data \cite{nfl2023big}, it is now possible to evaluate and better understand the offensive line positions in football at a fine-grained level.

\subsection{Contributions}

\cite{franks2015characterizing} used a Hidden Markov Model (HMM) to assign each defensive player to offensive players in professional basketball.  Their work demonstrated how player tracking data could be used to study defensive performance, an aspect of basketball that has historically received less attention than offensive performance. We apply a similar approach to football, where we use player tracking data to identify which defensive player each offensive lineman is blocking. In doing so, we seek to extend this type of analysis to another comparatively under-analyzed aspect of sports: offensive line performance in American football.

 For each offensive lineman, the model produces probabilistic blocking assignments associated with each frame of the tracking data. To our knowledge, no prior work generates blocking responsibilities as a frame-by-frame probabilistic quantity in continuous time. We show that these assignments offer useful summaries that can enhance existing analytical frameworks. Specifically, the assignment probabilities let us (i) measure the attention a rusher commands; (ii) augment adjusted plus-minus metrics \citep{Macdonald+2012} by assigning each blocker partial, continuous credit for the STRAIN \citep{nguyen2023here} of the rushers they engage in proportional to their assignment; (iii) read block shedding and disengagement directly off the assignment dynamics; and (iv) quantify how much space a rusher generates for their teammates. Further, we show that our methods are able to recover per-position blocker positioning, identify top pass rusher and blocker talent in accordance with scouting consensus, as well as recover hand-labeled pass-blocking assignments without the reliance on hand-labeled data. In the next Section, we describe the player tracking data and other derived metrics in detail.
In Section \ref{sec:methods}, we provide an overview of our methodology and present the main results.  Finally, we discuss potential future work and limitations in Section \ref{discussion}.

\section{Data} \label{data}
We use the player tracking data provided by the 2023 NFL Big Data Bowl \citep{howard2023nfl} in the analysis that follows.  This data comes from the NFL's Next Gen Stats system \citep{nfl2023ngs}, which returns the spatial coordinates of all 22 players on the field and the location of the football at a frequency of 10 Hz (i.e., 10 frames per second).  The data also includes player speed, acceleration, orientation, and event annotations associated with each time frame within a play.  Event annotations include time-stamped demarcations of actions that occurred during the play including, ball snap, quarterback throw, run, sack, etc. 

In total, the data contain 698 distinct pass rushers and 534 distinct pass blockers (i.e., offensive linemen, backs, and TEs in blocking roles) are analyzed across 8,530 plays during the first eight weeks of the 2021 season.  (Appendix~\ref{sec:positions} describes the offensive position groups these blockers are drawn from and how much each of them actually blocks on a pass play.) Throughout the paper, we index quantities by play $i$, pass rusher $j$, pass blocker $b$, and frame (time step) $t$, with frames sampled at $10$\,Hz. In order to assess pass blockers and pass rushers, we limit our analysis to the individuals tagged by Pro Football Focus (an independent NFL contractor providing additional data augmentation that relies on domain knowledge annotators) with a ``Pass Block'' or ``Pass Rush'' role. Furthermore, we focus on the time-frames between the snap and the quarterback event (i.e. pass attempt, run attempt, sack), again using the event annotations to filter our data. Before beginning our analysis, we smooth the raw positional data using a Kalman filter, the details of which are located in Appendix \ref{sec:kalman_filter}. 

In addition to the raw data, we generate a derived pass-rush metric called STRAIN \citep{nguyen2023here}, which is closely related to the concept of strain rate in materials science \citep{callister2018materials}.  Let $d_{ijt}$ be the Euclidean distance between rusher $j$ and the quarterback at frame $t$ of play $i$, and let $v_{ijt} = \mathrm{d}d_{ijt}/\mathrm{d}t$ be its rate of change. With this notation, \cite{nguyen2023here} define
\begin{equation}
\text{STRAIN}_{ijt} = - \frac{v_{ijt}}{d_{ijt}},
\end{equation}
as a per-frame measure of each rusher's pressure on the quarterback, so that a rusher's STRAIN is large when they are both close to and rapidly reducing their distance to the quarterback. Whereas \cite{nguyen2023here} estimate $v_{ijt}$ by a forward difference, we use the central difference method with second-order accuracy for a more numerically stable derivative, and likewise approximate the closing acceleration $a_{ijt} = \mathrm{d}v_{ijt}/\mathrm{d}t$ with a second-order central difference.

\section{A Hidden Markov Model (HMM) for Pass Blocking} \label{methods}
\label{sec:methods}

A Hidden Markov Model consists of a series of latent states that evolve in time and probabilistically ``emit'' observed data depending on the value of the latent state.  In our application, the latent states we seek to infer correspond to the identities of the rushers that a given offensive linemen could be blocking.  The conditional distribution of the location and orientation of the offensive linemen given the latent defensive matchup is called the ``emission distribution''. The blocking assignments evolve in time, as linemen can switch to block different rushers as the play progresses.  This probability of switching latent states (i.e., switching pass-blocking assignments) is called the ``transition probability".  Due to substantial differences in blocking technique and responsibility across offensive positions, the model is fit independently for each blocking position group $g \in \{\text{G},\text{T},\text{C},\text{TE},\text{RB},\text{WR},\text{FB}\}$. We write $g(b)$ for the position group of blocker $b$ and mark every group-specific parameter with the superscript $(g)$. 

\subsection{Modeling Blocker Positioning}

The core assumption of our model is that the pass-blocker maintain their position such that they are between their pass-rusher assignment and their quarterback. Let $R_{tj}$ denote the location of rusher $j$ and $Q_t$ the quarterback position, and define $O_{tb}$ as the location of blocker $b$ at time $t$. Let $I_{tbj}$ be an indicator variable equal to $1$ if blocker $b$ is assigned to rusher $j$ at frame $t$. Then, we posit a model for $O_{tb}$ given that they are matched up against rusher $j$ as
\begin{align}
\mu^{(g)}_{tj} &= \gamma^{(g)}_r R_{tj} + \gamma^{(g)}_q Q_t\\
O_{tb} \mid I_{tbj} = 1 &\sim \mathcal{N}_2\big(\mu^{(g(b))}_{tj},\; \sigma^{2(g(b))}_D\, \mathbf{I}_2\big)
\end{align}
where $O_{tb}, R_{tj}, Q_t \in \mathbb{R}^2$ are x-y field coordinates. The emission is a bivariate normal whose mean $\mu^{(g)}_{tj}$ is a point on the segment joining the pass rusher to the quarterback, with an isotropic covariance governed by  $\sigma^{2(g)}_D$. 

 We also model the orientation of blocker $b$ at time $t$ given that they are blocking rusher $j$. Since pass blockers typically face the pass rusher and attempt to stay square to the rusher during pass blocking events, we propose a model for the orientation of blocker $b$ based on the expectation that their orientation should be roughly opposite of that of his assigned rusher. Specifically, let $\eta_{tj}$ and $\eta_{tb}$ denote the orientation (in radians) of rusher $j$ and blocker $b$ at time $t$; the expectation that the blocker faces the rusher corresponds to $\eta_{tb} \approx \eta_{tj} + \pi$. Because orientation is a circular (angular) quantity, we model it directly with a von Mises distribution, the circular analogue of the Gaussian distribution.  Given that they are guarding rusher $j$, we assume blocker $b$'s orientation is $\text{vonMises}(\eta_{tj} + \pi, \, \kappa)$ with density
$$ p(\eta_{tb} \,|\, I_{tbj} = 1) = \frac{\exp\{\kappa \cos(\eta_{tb} - \eta_{tj} - \pi)\}}{2\pi I_0(\kappa)} = \frac{\exp\{-\kappa \cos(\eta_{tb} - \eta_{tj})\}}{2\pi I_0(\kappa)}, $$
where $I_0$ is the modified Bessel function of the first kind of order zero. The mean direction $\eta_{tj} + \pi$ encodes the blocker facing opposite the rusher, so the density is maximized when $\eta_{tb} = \eta_{tj} + \pi$. The fixed concentration $\kappa > 0$ controls how strongly orientation informs the assignment. We fix $\kappa = 1$. Because the positional and orientation emissions are conditionally independent given the assignment, we model the joint conditional emission density as the product of the two individual densities. In plain terms, the emission distribution encodes the idea that we expect a blocker to be facing their assignment as well as in between the assignment and the quarterback. 

\subsection{Modeling Switches}
In American football, offensive lineman may "pass off" rushers to their teammates, which results in a switch in assignment during the course of the play. Lineman may also "stunt" (feint as if they are going towards a rusher, then recover back to their original assignment), which also constitutes a change in assignment. Our model must encode this frame to frame transition, or "stickiness" of a blocker to his assignment.
 We let $n_i$ denote the number of pass rushers on play $i$, and we fit a separate  $\rho^{(g)}$ for each blocking position group $g$ (Appendix~\ref{sec:positions}):
$$ P(I_{tbj} = 1 | I_{(t-1)bj} = 1) = \rho^{(g(b))}, $$
$$ P(I_{tbj} = 1 | I_{(t-1)bj^{'}} = 1) = \frac{1 - \rho^{(g(b))}}{n_i -1},  \quad  \text{ for } j \neq j^{'}$$
Each $\rho^{(g)}$ is estimated via maximum likelihood within the EM algorithm, where it reduces to the expected fraction of retained assignments; the derivation is given in Appendix~\ref{sec:inference}.

The transition above treats every blocker in a position group as equally likely to switch at each frame. In practice, this transition probability varies by player: an elite tackle anchored on an edge rusher rarely switches, whereas an interior lineman routinely passes off twists and stunts. We therefore let each blocker's $b$ transition parameter $\rho_{b}$ vary by position and place a common Beta prior over the player-level transition probability parameter within each position group,
$$ \rho_{b} \sim \text{Beta}\big(\alpha^{(g(b))}_\rho, \beta^{(g(b))}_\rho\big). $$
The hyperparameters $(\alpha^{(g)}_\rho, \beta^{(g)}_\rho)$ are themselves estimated from the data via an empirical Bayes strategy, so that the player-level estimates are partially pooled. A blocker seen on many snaps is governed by his own transitions, while one seen rarely is shrunk toward the position-level mean $\alpha^{(g)}_\rho/(\alpha^{(g)}_\rho+\beta^{(g)}_\rho)$ (See Appendix \ref{sec:inference} for full details). This yields a regularized  measure of how readily each individual blocker switches or passes off assignments. Estimation of $\rho_{b}$ and $(\alpha^{(g)}_\rho,\beta^{(g)}_\rho)$ within the EM algorithm is detailed in the appendix.

We also incorporate pre-snap information to strategically inform the prior distribution of assignments. Before the snap, a blocker's responsibility is largely dictated by the protection scheme and the defensive alignment. For instance, a left tackle is a-priori likely to be responsible for the edge rusher aligned to its side, modulated by down, distance, and formation. Rather than initializing the hidden Markov chain with a uniform distribution over rushers, we encode this pre-snap belief with a multinomial-logistic model. Let $X_{ibj}$ denote a vector of pre-snap covariates for the (blocker $b$, rusher $j$) pair on play $i$ --- the relative alignment at the snap (e.g. lateral and longitudinal offset, whether the rusher is on the blocker's side and whether it is the outermost rusher on that side), the rusher's lined-up alignment, and interactions of these with down, distance, and formation. The initial distribution, omitting the play level subscript $i$ for notational brevity, is then
$$ P(I_{0 b j} = 1) = \frac{\exp\{\beta^{(g(b))\top} X_{0 b j}\}}{\sum_{j^{'}} \exp\{\beta^{(g(b))\top} X_{0b j^{'}}\}}, $$
with coefficients $\beta^{(g)}$ that are shared across plays but estimated separately for each blocking position group. Each position has their own learned mapping from pre-snap alignment to likely assignment, reflecting role-specific heterogeneity. The coefficients are fit within the EM algorithm (Appendix). 

 To appropriately account for the fact that not every pass-blocker needs to be assigned a rusher, we augment an $(n_i + 1)$-th ``disengaged state'' to each blocker's latent state space. Because the disengaged state does not correspond to a rusher, it has a different emission distribution. We assume that if a blocker is disengaged, their positional density is uniform over position and orientation on the field, so the disengaged state is assigned to blocker for a frame when no rusher's position-and-orientation density clears the background level. Transitions to and from the disengaged state are asymmetric, reflecting the fact that once a block is shed (denoted with parameter $p_{\text{fail}}$) or lost, it is difficult to re-engage (denoted with parameter $\rho^{(g)}_{\text{null}}$). As with $\rho_{b}$, the hazard $p_{\text{fail},b}$ is estimated per blocker, $\rho^{(g)}_{\text{null}}$ is pooled per position group, and the background level $c_{\text{bg}}$ is a single fixed constant shared across groups (See Appendix \ref{sec:inference} for details). The engaged-to-disengaged transition parameter is used in Section~\ref{shedding} to measure how long blockers sustain their blocks and how quickly rushers shed them.

To summarize the scope of each parameter: the emission weights $(\gamma^{(g)}_r, \gamma^{(g)}_q)$ and variance $\sigma^{2(g)}_D$, the pre-snap matchup coefficients $\beta^{(g)}$, the Dirichlet concentration $\alpha^{(g)}$ of the stay/switch/fail prior, and the null persistence $\rho^{(g)}_{\text{null}}$ are estimated separately for each blocking position group $g$; the stickiness $\rho_b$ and block-failure hazard $p_{\text{fail},b}$ are per blocker, partially pooled toward their position group; and the orientation concentration $\kappa$ and background level $c_{\text{bg}}$ are fixed constants common to all position groups. Each position group is fit on its own data. The assignment probabilities are recovered from the Expectation step of the forward-backward EM algorithm used to estimate the relevant parameters. A full treatment of algorithm details are given in the appendix, and code for the implementation can be found on \href{https://github.com/IshTheCoder/Matchup_Identifier}{Github}.

\subsection{Results of the HMM fits}

\begin{table}[ht]
\centering
\footnotesize
\begin{tabular}{|l|c|c|c|c|c|c|}
\hline
Position & $\gamma_r$ & $\gamma_q$ & $\sigma^2_{D}$ & $\rho$ & $p_{\text{fail}}$ & $\rho_{\text{null}}$ \\
\hline
Tackle        & 0.82 & 0.18 & 1.41 & 0.97 & 0.02 & 1.00 \\
Guard         & 0.84 & 0.16 & 1.21 & 0.97 & 0.01 & 1.00 \\
Center        & 0.84 & 0.16 & 1.05 & 0.98 & 0.01 & 0.97 \\
Tight End     & 0.89 & 0.11 & 2.29 & 0.96 & 0.02 & 1.00 \\
Fullback      & 0.79 & 0.21 & 4.57 & 0.70 & 0.27 & 1.00 \\
Running Back  & 0.55 & 0.45 & 5.91 & 0.39 & 0.58 & 1.00 \\
Wide Receiver & 0.99 & 0.01 & 4.61 & 0.45 & 0.55 & 1.00 \\
\hline
\end{tabular}
\caption{HMM parameter estimates by blocking position: the emission position weights $\gamma_r,\gamma_q$ (blocker location along the rusher--quarterback segment) and positional variance $\sigma^2_{D}$; the per-player stickiness $\rho$ (probability of staying on the engaged rusher); the block-failure  $p_{\text{fail}}$ (engaged $\to$ disengaged); and the disengaged-state persistence $\rho_{\text{null}}$.}
\label{tab:hmm_estimate}
\end{table}

We run the HMM for each position group and recover the emission weights $\gamma_r,\gamma_q$ which locate a blocker along the segment from his engaged rusher to the quarterback. Consistent with common protection designs, we find that offensive linemen and tight ends position themselves close to the rusher they engage with ($\gamma_r \approx 0.82$--$0.89$), whereas running backs position themselves closer to the quarterback ($\gamma_q \approx 0.45$). The positional variance $\sigma^2_{D}$ is tightest for the interior line ($\approx 1.0$--$1.4$) and largest for running backs, wide receivers, and fullbacks ($\approx 4.6$--$5.9$), who are typically not anchored to a fixed rusher.

The transition parameter, $\rho$,  is very high for linemen and tight ends ($\rho \approx 0.96$--$0.98$) indicating these positions rarely switch. It is far lower for running backs and wide receivers ($\rho \approx 0.4$), who are known to release into routes rather than sustaining a block over the course of the play. The structured disengaged state separates that release from a genuine switch through the block-failure parameter $p_{\text{fail}}$, the per-frame rate of transitioning from blocking to the disengaged state. It is negligible for the interior line and tight ends ($p_{\text{fail}} \approx 0.01$--$0.02$) and large for backs and receivers ($p_{\text{fail}} \approx 0.55$--$0.58$). Finally, the disengaged-state persistence $\rho_{\text{null}} \approx 0.97$--$1.00$ is near-absorbing at every position. Once a blocker is beaten or has released, he rarely re-engages within the same play. These estimates align closely with football intuition, supporting the model construction.

\subsubsection{Rusher Attention and Blocker Entropy}
\label{sec:attention}

In this Section, we describe simple metrics derived from the results of the HMM.  First, we derive a metric that quantifies the effective number of blockers a rusher faces at a given instant.  Specifically, we define ``attention'' as the sum of the assignment probabilities assigned to a rusher at a given time-step, $n_{\text{eff}}(j,t)=\sum_b \theta(b,j,t)$. Aggregated over a play or a season, it is the number of ``effective'' blockers a rusher faces over that time period. Because the number of effective blockers depends partly on the blocking scheme a defense faces, we normalize each rusher's attention within every play-frame relative to the average rusher on the field such that the resulting figures are comparable across three-, four-, and five-man fronts. Table~\ref{tab:att_top_norm} reports the rushers who draw the most front-normalized attention by position, which generate about 1.5 times as many blockers per play as the average rusher does. The five who draw the least by position can be found in Appendix Table~\ref{tab:att_bot_norm}.

\begin{table}[h!]
\centering
\begin{subtable}{0.24\textwidth}
\centering
\scriptsize
\begin{tabular}{lc}
\toprule
Name & Att. \\
\midrule
Adam Butler & 1.55 \\
Ta'Quon Graham & 1.49 \\
Calais Campbell & 1.47 \\
Dre'Mont Jones & 1.47 \\
Shelby Harris & 1.46 \\
\bottomrule
\end{tabular}
\caption{Edge}
\end{subtable}
\hfill
\begin{subtable}{0.24\textwidth}
\centering
\scriptsize
\begin{tabular}{lc}
\toprule
Name & Att. \\
\midrule
Marlon Davidson & 1.53 \\
Malcom Brown & 1.53 \\
Folorunso Fatukasi & 1.52 \\
Naquan Jones & 1.51 \\
Sheldon Richardson & 1.50 \\
\bottomrule
\end{tabular}
\caption{DT}
\end{subtable}
\hfill
\begin{subtable}{0.24\textwidth}
\centering
\scriptsize
\begin{tabular}{lc}
\toprule
Name & Att. \\
\midrule
Brandon Williams & 1.55 \\
Johnathan Hankins & 1.52 \\
Christian Covington & 1.51 \\
John Jenkins & 1.50 \\
Greg Gaines & 1.46 \\
\bottomrule
\end{tabular}
\caption{NT}
\end{subtable}
\caption{Top 5 pass rushers by front-normalized attention, by position (min 50 snaps).}
\label{tab:att_top_norm}
\end{table}

Just as attention summarizes the load a rusher draws, the assignment probabilities also characterize how a blocker distributes their responsibility. A blocker who stays attached to a single rusher concentrates all of their assignment mass on one rusher, whereas a blocker who passes off rushers, picks up stunts, or works in combination blocks spreads that mass across several. We quantify this with the entropy of a blocker's marginal assignment distribution over the rushers on a play, normalized by $\log$ of the number of rushers in the front so that it is comparable across fronts of different sizes. Table \ref{tab:blocker_entropy} ranks blockers by this front-normalized assignment entropy. The "stickiest" blockers (lowest entropy) are almost exclusively offensive tackles, who tend to man up on a single edge rusher; the one exception is a fullback, who blocks on too few snaps to spread his assignment. The most variable (highest entropy) are predominantly interior linemen, consistent with their role in passing off twists, stunts, and looping rushers.

\begin{table}[h!]
\centering
\begin{subtable}{0.48\textwidth}
\centering
\footnotesize
\begin{tabular}{llc}
\toprule
Name & Pos & Norm. Entropy \\
\midrule
Ben Cleveland & G & 0.368 \\
Netane Muti & G & 0.336 \\
Elijah Wilkinson & T & 0.308 \\
Michael Deiter & C & 0.307 \\
Calvin Throckmorton & T & 0.304 \\
Jamarco Jones & T & 0.303 \\
Justin Britt & C & 0.302 \\
Wes Schweitzer & G & 0.300 \\
Ben Bartch & T & 0.295 \\
Tyler Shatley & G & 0.292 \\
\bottomrule
\end{tabular}
\caption{Highest (most switching)}
\end{subtable}
\hfill
\begin{subtable}{0.48\textwidth}
\centering
\footnotesize
\begin{tabular}{llc}
\toprule
Name & Pos & Norm. Entropy \\
\midrule
Patrick Ricard & FB & 0.042 \\
Dion Dawkins & T & 0.134 \\
Isaiah Wynn & T & 0.140 \\
Yosuah Nijman & T & 0.147 \\
Rashawn Slater & T & 0.152 \\
Andrew Thomas & T & 0.154 \\
Donovan Smith & T & 0.158 \\
Dan Moore & T & 0.158 \\
Andrew Whitworth & T & 0.160 \\
Samuel Cosmi & T & 0.163 \\
\bottomrule
\end{tabular}
\caption{Lowest (stickiest)}
\end{subtable}
\hfill
\caption{Pass blockers ranked by front-normalized assignment entropy (min 50 snaps).}
\label{tab:blocker_entropy}
\end{table}

\subsubsection{Disengagement and Block Shedding} \label{shedding}
There are three distinct situations that can lead to blockers who will transition to a disengaged state: 1) a blocker can been be beaten, i.e. the the rusher has crossed to the quarterback side of him, 2) a blocker can win his matchup by driving the rusher away from the pocket or 3) a blocker can simply release the rusher with no other rusher to engage. In order to distinguish these different situations, we use geometric features together with STRAIN. We assume a blocker in a disengaged state is beaten only when the "assigned" rusher (i.e. the rusher whose assignment probability is highest amongst all possible rushers) is on quarterback-side of him.  Specifically, the blocker is considered beaten when  ($d(\text{rusher},\text{QB}) < d(\text{blocker},\text{QB})$), and that rusher's STRAIN is rising. Conversely, we assume the blocker has won his matchup when in a disengaged state and the "assigned" rusher is moving away with low or negative STRAIN. Finally, if neither of the previous two conditions are met, the frame is classified as a release.

For each play and blocker, we take the time from the snap to the first time beaten by a rusher as a survival outcome, right-censored at the throw for the majority of blocks that are never beaten, and summarize it with the Kaplan--Meier restricted mean survival time $\mathrm{RMST}_b$ alongside the per-block beat rate $\mathrm{BR}_b$. Computing the same two quantities for a rusher rather than a blocker gives $\mathrm{RMST}_j$ and the shed rate $\mathrm{BR}_j$, which measure how quickly and how often he beats the blocks he faces. Table~\ref{tab:block_engagement} ranks tackles by engagement. Because most blocks survive to the throw, the RMST sits near the $\sim\!4$\,s play horizon for nearly everyone. Elite pass protectors such as Rashawn Slater, Andrew Thomas, Trent Williams, and Lane Johnson are beaten on only $3$--$6\%$ of their blocks. Conversely, Table~\ref{tab:rusher_shedding} ranks rushers by how quickly they beat their blocks (shortest survival): Dont'a Hightower, Jonathan Greenard, Nick Bosa, and A.J.\ Epenesa beat blocks both fastest and most often, consistent with their reputations as disruptive rushers.

Since elite tackles also face the best rushers, we additionally report an opponent-adjusted quantity. We model the per-frame block-failure hazard directly as a discrete-time logistic survival on the beaten events,
$$ \operatorname{logit} h_{b,t} = \alpha + g_1\,\tilde t + g_2\,\tilde t^{2} + \beta_{\text{opp}}\,R_{j(b)} + u_b, $$
where $\tilde t$ is (standardized) frames since engagement, $R_{j}$ is the engaged rusher's peak-STRAIN plus-minus effect from the play-level model of Appendix~\ref{sec:playlevel}, entered at its posterior mean (standardized) as a fixed opponent-quality covariate, and $u_b$ is a regularized per-blocker hold frailty with scale $\sigma_u$. This model is also fit by NUTS with four chains ($1{,}000$ warmup and $1{,}000$ retained iterations each). The dwell coefficient is positive ($\hat{g}_1 > 0$) suggesting that blocks fail more readily the longer they are asked to last. The opponent coefficient is positive too ($\hat{\beta}_{\text{opp}} > 0$), confirming that facing a better rusher raises the failure hazard. The resulting hold rating $-u_b$ (now adjusted for opponent quality) moves Trent Williams from $15$th to $3$rd among the $93$ tackles with at least $40$ blocks (Figure~\ref{fig:block_hold}). The implied per-play beat probabilities run from $\sim\!3\%$ for the best protectors to $\sim\!20\%$ for the most-beaten, rising a point or two against an elite edge rusher.

\section{Extending Plus Minus Models} \label{plusminus}

In this section, we describe how we can utilize partial blocking responsibilities from the HMM assignment probabilities to quantify the moderated effect of a pass blocker on a rusher's STRAIN towards the quarterback. We introduce the continuous generalization which operates on a frame-by-frame basis. The continuous generalization allows us to model the change in STRAIN over time-steps attributed to a rusher and his blocking counterparts. We rank the rushers and blockers according to expected STRAIN generated (or prevented), and find that our measures identifies talent that correlates with PFF's hand tagged rusher and blocker grades.  Our model presented here is Bayesian, allowing us to quantify the uncertainty of the estimates of our parameters of interest. 

\subsection{Continuous Time Blocker Plus-Minus }\label{continuous}

We extend the play-level plus-minus model of Appendix~\ref{sec:playlevel} to continuous time by resolving it to the (play $i$, rusher $j$, frame $t$) level, so that every rusher-frame is a distinct observation rather than a single play-level summary. We take the one-frame change in STRAIN as the response, $\Delta\text{STRAIN}_{ijt} = \text{STRAIN}_{i,j,t+1} - \text{STRAIN}_{ijt}$. The regressor is the blocking effect a rusher receives at $t$, weighted by the blocking responsibility $\theta(b,j,t)$. Because the response looks one frame ahead, $\theta(b,j,t,i)$ is the filtered assignment posterior, which uses the tracking data only up to frame $t$. The blocker effect $B^{\Delta}_b$ reflects how much blocker $b$'s engagement decelerates the rusher over the next frame.
$$ \Delta\text{STRAIN}_{ijt} \sim \mathcal{N}\Big( \mu + R^{\Delta}_{j} + Q^{\Delta}_{q_i} - \!\!\sum_{b \in \mathcal{X}_i}\!\! B^{\Delta}_{b}\,\theta(b, j, t, i) + \phi\,\text{STRAIN}_{ijt},\; \sigma_{\Delta}^2 \Big). $$
Equivalently, the model is a first-order autoregression of the next-frame STRAIN with coefficient $1+\phi$. $R^{\Delta}_j$ and $Q^{\Delta}_{q_i}$ are constant rusher and quarterback random intercepts that capture a rusher's own drive and a quarterback's pocket behaviour. The rusher, blocker, and quarterback effects are drawn around a linear function of player attributes $z = [\text{position}, \text{height}, \text{weight}]$, which shrinks each player toward their positional/size archetype,
$$ R^{\Delta}_j = \alpha_R^{\Delta\top} z_j + \sigma^{\Delta}_R\,\varepsilon^R_j, \quad B^{\Delta}_b = \alpha_B^{\Delta\top} z_b + \sigma^{\Delta}_B\,\varepsilon^B_b, \quad Q^{\Delta}_q = \alpha_Q^{\Delta\top} z_q + \sigma^{\Delta}_Q\,\varepsilon^Q_q, \qquad \varepsilon \sim \mathcal{N}(0,1), $$
with their own scales and residual $\sigma_{\Delta}$. The scales are given half-normal priors, $\sigma^{\Delta}_B \sim \mathcal{N}^+(0, 0.05^2)$ and $\sigma^{\Delta}_R, \sigma^{\Delta}_Q \sim \mathcal{N}^+(0, 0.02^2)$; $\sigma_{\Delta} \sim \mathcal{N}^+(0, 0.5^2)$. The intercept $\mu$, the coefficient $\phi$, and the attribute coefficients $\alpha^{\Delta}_R, \alpha^{\Delta}_B, \alpha^{\Delta}_Q$ receive $\mathcal{N}(0,1)$ priors.

\begin{figure}[h!]
\centering
\includegraphics[width=0.72\textwidth]{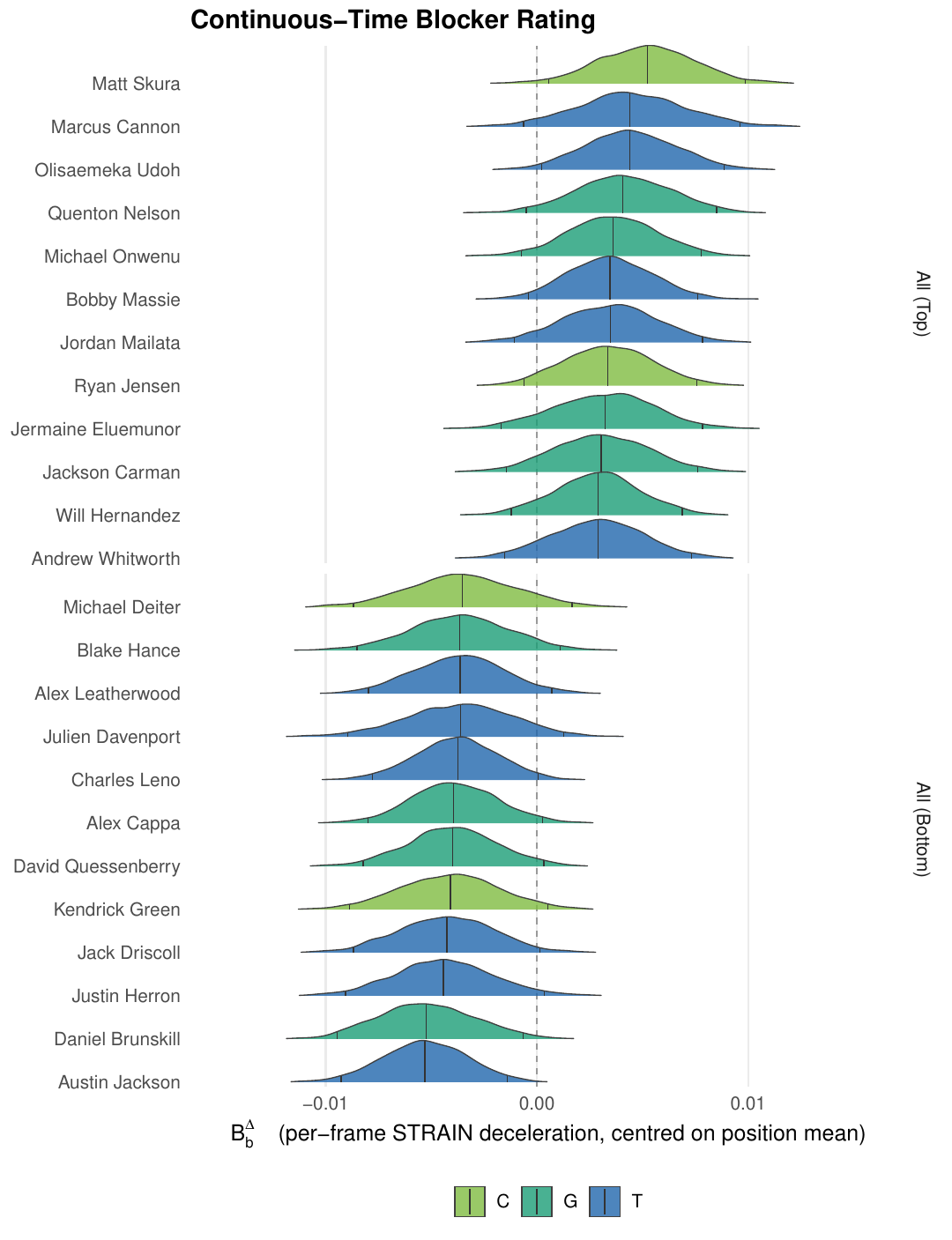}
\caption{Posterior distributions of the continuous-time blocker rating $B^{\Delta}_b$ for the top and bottom twelve pass blockers ($\geq 100$ snaps), coloured by position. Vertical lines mark the $2.5$, $50$ and $97.5$ percentiles. Zero is the average player at the position. Like the play-level coefficient, this is an effect per unit of assignment.}
\label{fig:blocker_continuous}
\end{figure}

The continuous-time model gives rusher, blocker, and quarterback ratings derived from STRAIN dynamics. We fit it by NUTS on the $1.1$M rusher-frames over eight weeks, again with four chains ($800$ warmup and $800$ retained iterations each, $3{,}200$ draws); the mean-reversion coefficient is $\hat{\phi} = -0.059$. On the common set of $207$ linemen with at least $50$ PFF pass-block snaps, the per-blocker effect $B^{\Delta}_b$, for which higher values mean better protection, correlates negatively with independent PFF charting on both pressures allowed (Pearson $-0.32$) and the beaten rate ($-0.25$), and it also separates protection skill within position (Pearson $-0.29$). The rankings (Figure~\ref{fig:blocker_continuous}) are led by interior linemen Matt Skura, Quenton Nelson, and Michael Onwenu alongside tackles Marcus Cannon, Olisaemeka Udoh, Bobby Massie, and Jordan Mailata. The rusher effect $R^{\Delta}_j$ in turn tracks charted pass-rush production closely (Section~\ref{pffvalid}).

\begin{figure}[h!]
\centering
\includegraphics[width=0.85\textwidth]{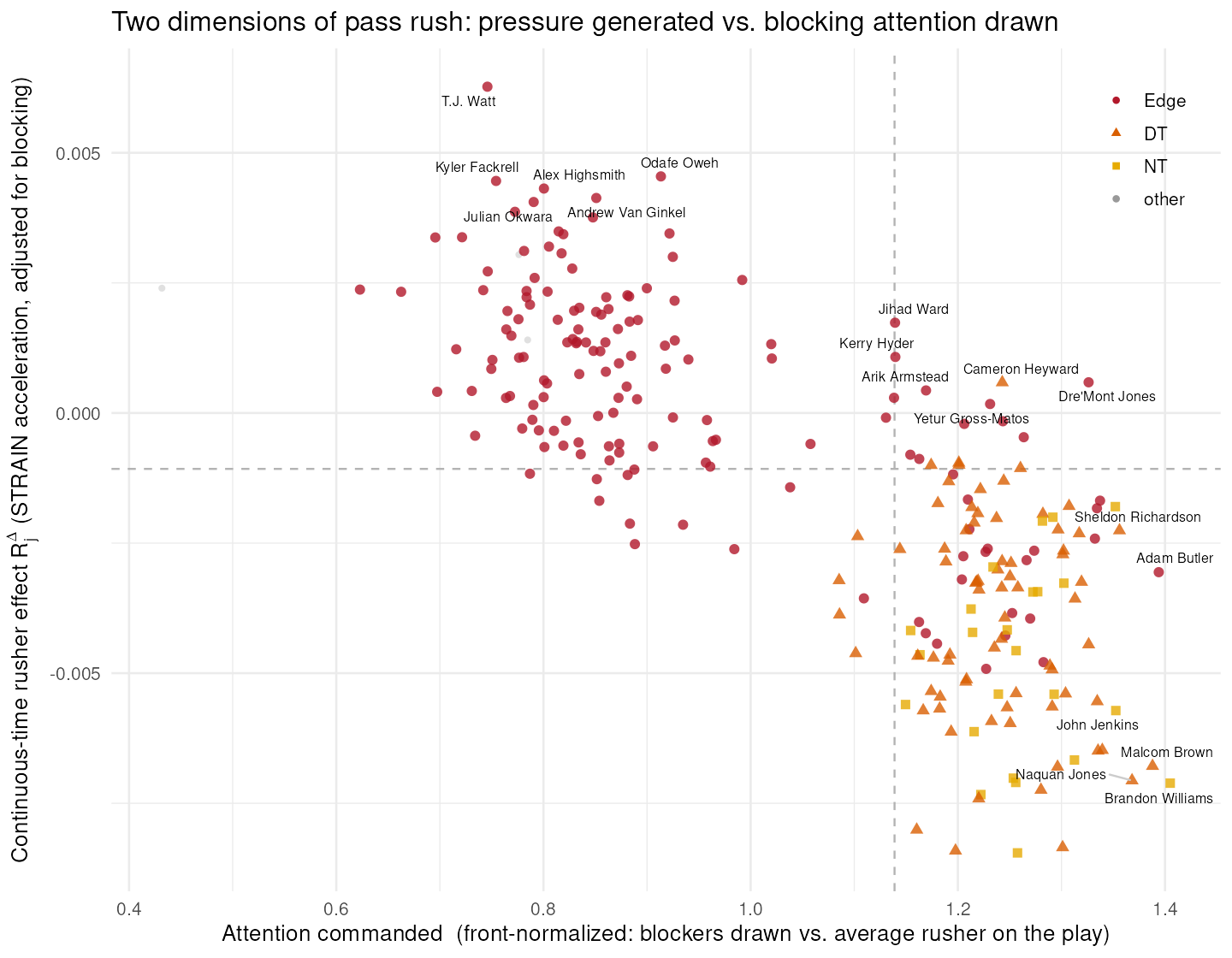}
\caption{The two dimensions of pass-rush value for every rusher with $\geq 50$ snaps ($n=250$). The horizontal axis is front-normalized attention commanded; the vertical axis is the continuous-time rusher effect $R^{\Delta}_j$ (per-frame STRAIN acceleration, adjusted for the blocking faced).}
\label{fig:rusher_2d}
\end{figure}

Combining the attention metric of Section~\ref{sec:attention} with the continuous-time rusher effect, we can plot how quickly a rusher builds STRAIN against how much blocking attention he receives. We quantify attention in front-normalized terms: for each snap we divide a rusher's effective-blocker count $\sum_b \theta(b,j,i)$ by the average over the rushers active on that snap, then average over the rusher's snaps, so a value of $1$ is an average share and values above $1$ mean the rusher draws more than his teammates. Figure~\ref{fig:rusher_2d} plots the two dimensions for every rusher with at least fifty pass-rush snaps. The two are strongly negatively related (Pearson $-0.78$), and the plot splits positionally. Edge rushers, who typically face fewer blockers, command less attention and sit in the upper left, led by T.J.\ Watt, Odafe Oweh, Alex Highsmith, and Kyler Fackrell. Interior linemen draw more blockers but build STRAIN more slowly and sit in the lower right (e.g.\ Brandon Williams, Malcom Brown, and Naquan Jones). The few rushers in the upper right, who draw extra attention and still build STRAIN quickly, include Jihad Ward, Kerry Hyder, Arik Armstead, and Dre'Mont Jones, along with Cameron Heyward among interior linemen.

\section{Pocket Space Ownership} \label{pocketspace}
\begin{figure}[h!]
\centering
\includegraphics[width=\textwidth]{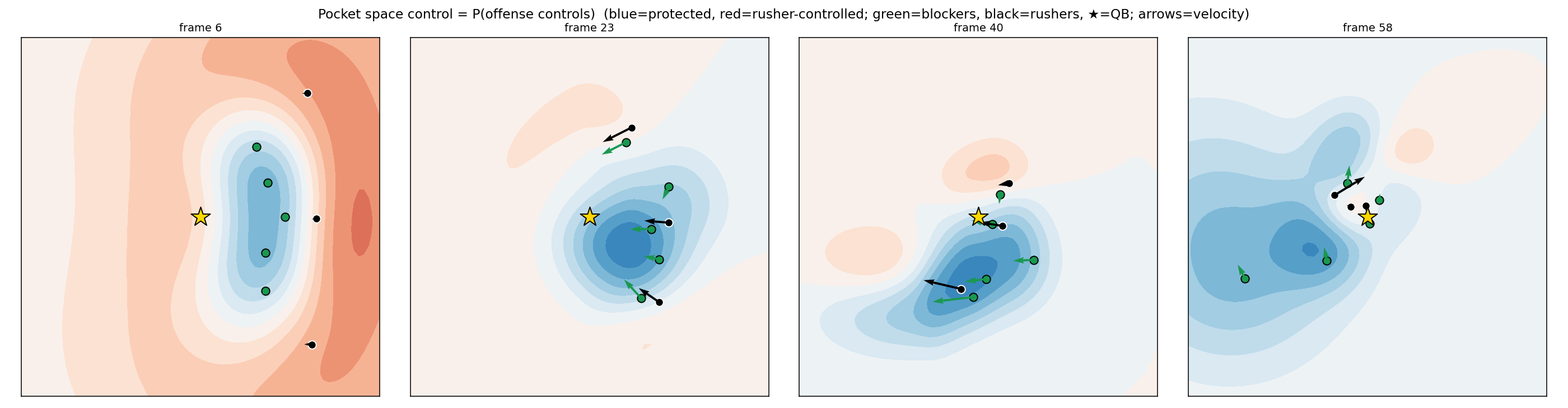}
\caption{Pocket space control across a collapsing play (snap to sack, left to right). Color is the probability that the offense (blockers) controls each location; the quarterback ($\star$) sits in a protected blue pocket early and is exposed as a rusher wins control. Green: blockers; black: rushers; arrows: velocity.}
\label{fig:pocket_control}
\end{figure}

Our final application incorporates the assignment probabilities into a space-control model of the pass-protection pocket. We borrow heavily from \cite{fernandez2018wide} a space control approach application in soccer and utilize the fitted blocking probabilities to measure the space a rusher generates for his teammates. Each player $k$ is assigned a bivariate influence $I_k(p,t)\in(0,1]$ over field locations $p$. A location is controlled by whichever side's influence is greater. Writing $\mathcal{B}_t$ for the blockers (including the quarterback) and $\mathcal{R}_t$ for the rushers on the field at time $t$, the pocket control (PC) at location $p$ is the probability the offense owns that space,
\begin{equation}
\mathrm{PC}(p,t) \;=\; \mathrm{logit}^{-1}\Big( \textstyle\sum_{b \in \mathcal{B}_t} I_b(p,t) \;-\; \sum_{j \in \mathcal{R}_t} I_j(p,t) \Big) \label{eq:pc}
\end{equation}
visualized in Figure~\ref{fig:pocket_control}. Figure~\ref{fig:pocket_control} shows the field across a collapsing pocket: at the snap the quarterback sits in a protected area, which shrinks as a rusher wins. Quantitatively, the minimum over a play of the quarterback's own controlled area predicts charted pressure and sacks. 
To attribute that control to individual players rather than to a side, we also define each player's control share  as $w_k(p,t) = I_k(p,t) / \sum_{k'} I_{k'}(p,t)$. Since space near the quarterback is more important, we value it with weight $V(p)=e^{-\lVert p-\mathrm{QB}_t\rVert/\lambda}$, which decays exponentially with distance $\lambda = 3$\,yd from the quarterback.  Integrating the two over a disk $\mathcal{D}_t$ of radius $5$\,yd centred on the quarterback gives each $k$ a value-weighted owned-space quality
\begin{equation}
Q_k(t) \;=\; \sum_{p \in \mathcal{D}_t} w_k(p,t)\, V(p)\, \Delta a , \label{eq:q}
\end{equation}
evaluated on a quarterback-centred grid, where $\Delta a$ is grid partition's area. Table \ref{tab:pocket_rusher_space} lists rushers who have gained valuable access to the pocket. 

While ownership is valuable in identifying who is gaining ground near the quarterback, pass rushers rarely work in isolation. Typically, space generation occurs when a rusher pulls a blocker off a teammate, freeing that teammate to win valuable near-QB ground. Our HMM reports, as a fitted probability $\theta(b,r,t)\in[0,1]$, whether the blocker is blocking him. Over a short window $[t,t{+}w]$ we therefore weight the event ``blocker $b$ leaves teammate $i'$ for the generator $i$'' by the probabilities directly,
\begin{equation*}
W_b(i'\!\to\! i)=\underbrace{\theta(b,i';t)}_{\text{was on }i'}\,\underbrace{\big(1-\theta(b,i';t{+}w)\big)}_{\text{left }i'}\,\underbrace{\theta(b,i;t{+}w)}_{\text{now on }i}\,\underbrace{\big(1-\theta(b,i;t)\big)}_{\text{not on }i\text{ before}},
\end{equation*}
and credit the generator with that fraction of the value-weighted space his freed teammate then wins,
\begin{equation*}
\mathrm{SGG}_i \;\mathrel{+}=\; \sum_{b}\sum_{i'\neq i} W_b(i'\!\to\! i)\,\big(Q_{i'}(t{+}w)-Q_{i'}(t)\big).
\end{equation*}

This recovers a dimension of pass rushing that personal-pressure metrics miss. The leading space generators (Appendix Table~\ref{tab:pocket_sgg}) are primarily interior linemen  who command double teams and, in doing so, free their teammates. They rank high in attention but low in personal pressure and plus-minus. \citet{fernandez2018wide} pair generation with its mirror image, space received. Summing the directed flow over each generator--receiver pair of teammates (Table~\ref{tab:pocket_pair_flow}) recovers recognizable synergies --- Aaron Donald drawing the double and freeing his nose tackle, Dalvin Tomlinson springing Danielle Hunter off the edge, Maxx Crosby opening the interior.

\begin{table}[h!]
\centering
\footnotesize
\begin{tabular}{lllc}
\toprule
Unit & Generator & Receiver & Space \\
\midrule
IND & Grover Stewart (DT) & DeForest Buckner (DT) & 98.4 \\
GB & Kingsley Keke (Edge) & Kenny Clark (NT) & 73.1 \\
MIN & Dalvin Tomlinson (DT) & Danielle Hunter (Edge) & 69.9 \\
LA & Aaron Donald (DT) & Sebastian Joseph (NT) & 59.7 \\
NYJ & Folorunso Fatukasi (DT) & John Franklin-Myers (Edge) & 51.4 \\
LV & Maxx Crosby (Edge) & Quinton Jefferson (DT) & 51.3 \\
DEN & Von Miller (Edge) & DreMont Jones (Edge) & 51.0 \\
DEN & Malik Reed (Edge) & DreMont Jones (Edge) & 47.8 \\
NYG & Austin Johnson (NT) & Leonard Williams (DT) & 46.2 \\
WAS & Jonathan Allen (DT) & Daron Payne (DT) & 46.2 \\
IND & DeForest Buckner (DT) & Grover Stewart (DT) & 46.0 \\
MIA & Adam Butler (Edge) & Christian Wilkins (DT) & 45.5 \\
\bottomrule
\end{tabular}
\caption{The strongest within-front space-generation flows: for each pair of teammates, the value-weighted near-QB space one rusher (the \emph{generator}) frees for the other (the \emph{receiver}) by drawing a blocker's assignment away, summed over eight weeks.}
\label{tab:pocket_pair_flow}
\end{table}

\section{External Validation against PFF Charting} \label{pffvalid}
Our assignments and metrics are derived purely from tracking geometry and STRAIN, so the independently hand-charted pass-block / pass-rush data from Pro Football Focus provides an external check.

First, the HMM assignments recover the charted blocker responsibilities. PFF tags, for each play and blocker, the single rusher that blocker was responsible for. Over the $44{,}037$ charted blocker-plays we are able to match, our smoothed assignment places a mean of $0.77$ of its engaged probability mass on exactly that rusher and the model's most-assigned rusher agrees with PFF's designation on $87.5\%$ of blocker-plays. 

Second, the block-shedding survival rates (Section~\ref{shedding}) and the continuous-time rusher and blocker effects (Section~\ref{continuous}) also correlate with charted pressure outcomes in the expected directions (Table~\ref{tab:pff_validation}). The table orients every correlation so that agreement in the expected direction is positive. On the rusher side the agreement is strong: a rusher's continuous-time effect $R^{\Delta}_j$ tracks his PFF pressure rate (Pearson $0.57$) and sack rate ($0.46$), and his shed rate tracks his pressure rate as well ($0.37$). On the blocker side the survival correlations are modest, but statistically significant and directionally correct: blockers who sustain their blocks longer allow fewer pressures ($0.19$ after orientation), and blockers who are beaten more often allow more ($0.17$). The continuous-time blocker rating $B^{\Delta}_b$ is the strongest blocker-side metric, with an oriented correlation of $0.32$ with pressures allowed. In conclusion, the rush-side metrics validate strongly against independent charting and the block-side metrics validate directionally, providing confidence that our methodology is picking up relevant signal.

\begin{table}[h!]
\centering
\footnotesize
\begin{tabular}{llcccc}
\toprule
Model metric & Symbol & PFF outcome & Sign $s$ & Pearson $\times s$ & Spearman $\times s$ \\
\midrule
\multicolumn{6}{l}{\emph{Pass rushers (pressures obtained)}} \\
\quad Continuous-time effect & $R^{\Delta}_j$ & pressures & $+1$ & $+0.57$ & $+0.58$ \\
\quad Continuous-time effect & $R^{\Delta}_j$ & sacks & $+1$ & $+0.46$ & $+0.46$ \\
\quad Shed rate & $\mathrm{BR}_j$ & pressures & $+1$ & $+0.37$ & $+0.36$ \\
\quad Engagement RMST & $\mathrm{RMST}_j$ & pressures & $-1$ & $+0.31$ & $+0.36$ \\
\midrule
\multicolumn{6}{l}{\emph{Pass blockers (pressures allowed)}} \\
\quad Continuous-time effect & $B^{\Delta}_b$ & pressures all. & $-1$ & $+0.32$ & $+0.32$ \\
\quad Beat rate & $\mathrm{BR}_b$ & pressures all. & $+1$ & $+0.17$ & $+0.31$ \\
\quad Engagement RMST & $\mathrm{RMST}_b$ & pressures all. & $-1$ & $+0.19$ & $+0.33$ \\
\bottomrule
\end{tabular}
\caption{Correlation of tracking-derived model metrics with independent PFF pass-pressure charting (players with $\geq 50$ PFF snaps; rushers $n=250$, blockers $n=207$). The Symbol column gives the quantity being correlated: the continuous-time rusher and blocker effects $R^{\Delta}_j$ and $B^{\Delta}_b$ of Section~\ref{continuous}, and the survival summaries $\mathrm{RMST}$ and beat rate $\mathrm{BR}$ of Section~\ref{shedding}, subscripted $j$ for a rusher and $b$ for a blocker. PFF outcomes are per-snap rates, where a pressure is a sack, hit, or hurry. Each correlation is multiplied by the sign $s$, so that a positive value means agreement with PFF in the expected direction. $s=-1$ for metrics that should move against the PFF outcome: a rusher whose blocks last longer (larger $\mathrm{RMST}_j$) should generate less pressure, and a blocker who decelerates the rush more (larger $B^{\Delta}_b$) or sustains his blocks longer (larger $\mathrm{RMST}_b$) should allow less. }
\label{tab:pff_validation}
\end{table}

\clearpage

\section{Discussion} \label{discussion}
We adapted the defensive-matchup hidden Markov model of \cite{franks2015characterizing} from basketball to football pass protection. The recovered assignments reproduce independently, hand-charted blocker responsibilities. Beyond individual evaluation, the assignment probabilities reveal which rushers draw the most interference and how protection schemes resolve over the course of a play, information that can inform opponent scouting and the design of blocking schemes. We show that integrating the probabilistic assignments into existing analytical frameworks yields insight into identifying top rusher and blocker talent across a wide array of metrics ranging from space conceded / generated to STRAIN allowed / prevented. Our implementation itself provides closed form updates where applicable, providing scalable inference and a framework for incorporating more complex modeling decisions. 

Although our methods are promising, there is certainly room for improvement. First, additional information could sharpen the HMM assignments. We rely primarily on positioning, yet a blocker's direction of velocity carry complementary signal. We also do not fully separate a blocker who is actively impeding a rusher from one who merely happens to be in the right place (i.e. a missed block while nominally ``assigned''). The disengaged state is a first step toward this distinction, but a two-layer HMM in the spirit of \cite{twolayerhmm} could, conditional on the assignment, additionally infer the engagement status. Second, blocker evaluation remains the hardest open problem. Our continuous time model is a step in the right direction; however, it is nonetheless an associational estimate. Reformulating the problem from a causal perspective could potentially provide more actionable insight for coaching and front office perspectives. Finally, other response variables are worth exploring. Rather than STRAIN, one could rate rushers by acceleration projected along the quarterback's escape direction, foregrounding pursuit ability, or model the time-to-pressure hazard directly. More broadly, the probabilistic instrument is agnostic to the outcome, so any frame-level measure of pressure or disruption could be substituted for STRAIN without changing the evaluation machinery.

\clearpage
\begin{appendices}
\section{Offensive Position Groups}
\label{sec:positions}
Our models are fit separately for each of the seven offensive position groups that appear in a blocking role: guard (G), tackle (T), center (C), tight end (TE), running back (RB), wide receiver (WR), and fullback (FB). These groups differ both in how often they block on a pass play and how they block.

The offensive linemen (tackles, guards, and centers) are the primary pass blockers. They start on the line of scrimmage on every snap and block on essentially every dropback. The center and the guards form the interior of the line and typically engage interior rushers, who attack on the short, direct path to the quarterback. The two tackles align at the outside edges of the line and usually handle edge rushers, who take the widest and longest path to the quarterback.

Tight ends occupy an intermediate role. A tight end may align at the end of the line or in the backfield, and on a given pass play he may either stay in to block or release into a route as a receiver. 

Running backs, wide receivers, and fullbacks are primarily receivers on pass plays; They do block on occasion. A running back or fullback may stay in to protect the quarterback, chip an edge rusher on his way into a route, or pick up an unblocked blitzing defender, and, in maximum-protection schemes, additional players are kept in to block. However, on most dropbacks these players release downfield rather than engage a rusher.

\begin{table}[h!]
\centering
\footnotesize
\begin{tabular}{lcc}
\toprule
Position group & Dropbacks with $\geq\!1$ blocking & Mean blockers per dropback \\
\midrule
Tackle (T)        & $8{,}530$ \ ($100.0\%$) & $2.09$ \\
Guard (G)         & $8{,}471$ \ ($99.3\%$)  & $1.93$ \\
Center (C)        & $7{,}820$ \ ($91.7\%$)  & $0.98$ \\
Running back (RB) & $1{,}541$ \ ($18.1\%$)  & $0.18$ \\
Tight end (TE)    & $1{,}195$ \ ($14.0\%$)  & $0.17$ \\
Fullback (FB)     & $146$ \ ($1.7\%$)       & $0.02$ \\
Wide receiver (WR)& $44$ \ ($0.5\%$)        & $0.01$ \\
\bottomrule
\end{tabular}
\caption{How much each offensive position group actually pass-blocks, over the $8{,}530$ dropbacks in the sample. A player counts as blocking on a play when the HMM places non-zero engaged assignment mass on him.}
\label{tab:posgroups}
\end{table}

\section{Smoothing Positional Data}
\label{sec:kalman_filter}
Before beginning our analysis using raw (x,y) tracking data, we smooth the observations. It is known that RFID sensors are noisy, and thus analysis of raw values are subject to error. More importantly, since our analysis heavily utilizes numerical differentiation techniques which rely on smooth function assumptions, de-noising the presented data is of significant importance.  In order to achieve this, we apply a Kalman filter to the positional data, using the reported velocity and acceleration only to initialize the latent state, as described below.

Noting that movement in the x and y direction can be calculated independently, we present the method for the x direction of motion and note that the same method follows for the y component.
\subsection{Kalman Filter Implementation}
A Kalman Filter is a specific instance of general state space models where both the emission and latent state distributions are assumed to be normally distributed. The general formulation is as follows:
$$  y_t \sim \mathcal{N}(F \theta_t, V) $$
$$  \theta_t \sim \mathcal{N}(G \theta_{t-1}, W) $$
$$  x_0 \sim \mathcal{N}(m_0, C_0)$$
Note that $F$ and $G$ are known matrices, and $V$ and $W$ represent the error associated with the observation vector $y_t$ and the latent state $\theta_t$ respectively. Similarly, $m_0$ and $C_0$ are initial parameters defining the distribution of the initial latent state $x_0$. 

In our case the latent state collects the player's true position, velocity, and acceleration in one coordinate,
$$ \theta_t = \begin{bmatrix} z_t \\ \dot{z}_t \\ \ddot{z}_t \end{bmatrix}, $$
with $z_t$ the latent position. The dynamics follow a constant-acceleration model over the frame interval $\delta = 0.1$\,s,
$$ z_t = z_{t-1} + \delta \dot{z}_{t-1} + \tfrac{1}{2}\delta^2 \ddot{z}_{t-1} + \tfrac{1}{6}\delta^3 j_t, \quad \dot{z}_{t} = \dot{z}_{t-1} + \delta\ddot{z}_{t-1} + \tfrac{1}{2}\delta^2 j_t, \quad \ddot{z}_{t} = \ddot{z}_{t-1} + \delta j_t, $$
that is $\theta_t = G\,\theta_{t-1} + j_t$ with
$$ G = \begin{bmatrix} 1 & \delta & \tfrac{1}{2}\delta^2 \\ 0 & 1 & \delta \\ 0 & 0 & 1 \end{bmatrix}. $$
The jerk $j_t$ is an unknown continuous white-noise input of spectral density $q$, which induces the standard full-rank process covariance
$$ W = q \begin{bmatrix} \delta^5/20 & \delta^4/8 & \delta^3/6 \\ \delta^4/8 & \delta^3/3 & \delta^2/2 \\ \delta^3/6 & \delta^2/2 & \delta \end{bmatrix}. $$

We treat position as the only observation and let the filter infer velocity and acceleration:
$$ y_t = z_t + \varepsilon_t, \qquad \varepsilon_t \sim \mathcal{N}(0, \sigma_p^2), \qquad F = \begin{bmatrix} 1 & 0 & 0 \end{bmatrix}. $$
The tracking feed reports a speed $\lVert v_t\rVert$, an acceleration magnitude $\lVert a_t\rVert$, and a direction of motion $\eta_t$. We use those derived kinematics to seed the initial state mean $\theta_0$, with a tight prior on position and a loose one on the inferred velocity and acceleration, $C_0 = \mathrm{diag}(0.01,\,1,\,4)$.

We set the process spectral density $q = 1~\text{yd}^2/\text{s}^5$ and the observation variance $\sigma_p^2 = 10^{-3}~\text{yd}^2$ $\big(\!\approx (0.03~\text{yd})^2\big)$. Finally, we apply the Kalman smoother, as derived in \cite{Kalman} and implemented in \cite{pykalman}, to every player, play trajectory, and use the smoothed position, velocity, and acceleration in all subsequent analysis.

\section{HMM Parameter Estimation}
\subsection{Model Formulation}
We model the evolution of pass blocking assignments, as given by the matrix of matchups, $I_i$,  over the course of a play $i$ using a hidden Markov model. Because the model is fit independently for each blocking position group, the derivation below describes a single group, and we suppress the group superscript $(g)$ throughout this appendix, reserving the superscript $(s)$ for the EM iteration. Every parameter estimated here is understood to be that group's own.

Each blocker $b$ on play $i$ is associated with $n_i + 1$ hidden states: the $n_i$ rushers on the play, and a $(n_i+1)$-th disengaged state. We write $I_{tbj} = 1$ when blocker $b$ occupies state $j \in \{1, \dots, n_i + 1\}$ at frame $t$, and reserve the index $j \le n_i$ for engaged states.

\paragraph{Emission Distribution} Removing the subscript $i$ for clarity, an engaged state emits the blocker's position and orientation as the product of the two following densities
$$ D_{tb} \mid I_{tbj} = 1, \gamma_q, \gamma_r, \sigma^2_D \sim \mathcal{N}_2\big(\gamma_r R_{tj} + \gamma_q Q_t,\; \sigma^2_D\,\mathbf{I}_2\big), \qquad j \le n_i, $$
$$ \eta_{tb} | I_{tbj} = 1, \eta_{tj} \sim \text{vonMises}\big(\eta_{tj} + \pi, \, \kappa\big), \qquad j \le n_i, $$
where $\kappa$ is a fixed hyper-parameter and the emission weights are constrained to the simplex, $\gamma_r + \gamma_q = 1$. The disengaged state has no rusher to correspond to, so we assign it an emission of a flat positional background and a uniform orientation,
$$ p(D_{tb} \mid I_{tb,n_i+1} = 1) = \exp(c_{\text{bg}}), \qquad p(\eta_{tb} \mid I_{tb,n_i+1} = 1) = \frac{1}{2\pi}, $$
with $c_{\text{bg}}$ a fixed constant (von Mises with $\kappa = 0$ for the orientation). The value of $c_{\text{bg}}$ is chosen by identifying the elbow of the implied mean disengaged probability curve (about $8\%$ of blocker-frames) on one week of data.

\paragraph{Transition Dynamics} The transition is structured so that disengaging is a distinct event from switching rushers. From an engaged state, blocker $b$ stays with probability $\rho_b$, moves to one of the other rushers with total probability $1 - \rho_b - p_{\text{fail},b}$ , disengages with the block-failure hazard $p_{\text{fail},b}$. Once in the disengaged state, a blocker stays in the disengaged state with probability $\rho_{\text{null}}$ and re-engages a rusher with $1 - \rho_{\text{null}}$:
\begin{align*}
P(I_{tbj} = 1 \mid I_{(t-1)bj} = 1) &= \rho_b, & j &\le n_i, \\
P(I_{tbj} = 1 \mid I_{(t-1)bj'} = 1) &= \frac{1 - \rho_b - p_{\text{fail},b}}{n_i - 1}, & j \ne j' &\le n_i, \\
P(I_{tb,n_i+1} = 1 \mid I_{(t-1)bj} = 1) &= p_{\text{fail},b}, & j &\le n_i, \\
P(I_{tb,n_i+1} = 1 \mid I_{(t-1)b,n_i+1} = 1) &= \rho_{\text{null}}, \\
P(I_{tbj} = 1 \mid I_{(t-1)b,n_i+1} = 1) &= \frac{1 - \rho_{\text{null}}}{n_i}, & j &\le n_i.
\end{align*}
Here $n_i$ is the number of rushers on play $i$, constant within a play; $\rho_b$ and $p_{\text{fail},b}$ are per blocker and $\rho_{\text{null}}$ is shared across the position group.

\paragraph{Initial Assignments} At the snap the engaged states share the conditional-logit matchup prior of Section~\ref{methods} and the disengaged state receives a fixed mass $\nu$,
$$ P(I_{0bj} = 1) = (1 - \nu)\,\frac{\exp\{\beta^\top X_{0bj}\}}{\sum_{j' \le n_i} \exp\{\beta^\top X_{0bj'}\}}, \quad j \le n_i, \qquad P(I_{0b,n_i+1} = 1) = \nu, $$
with $\nu = 0.05$. We omit the subscript for the play $i$ for notational brevity. 

\paragraph{Likelihood} Writing $\mathbf{I}$ for the matrix of assignments, $\mathbf{D}$ for the blocker positions, and $\boldsymbol{\eta}$ for the blocker orientations over a play, the complete-data likelihood over all plays in the group is
\[
\resizebox{\textwidth}{!}{$\displaystyle L = \prod_{i=1}^{N} \prod_{t,b} \prod_{j=1}^{n_i+1} \big[ P(D_{t_ib_i} \mid I_{t_i b_i j_i}) \, P(\eta_{t_ib_i} \mid I_{t_ib_ij_i}) \, P(I_{t_ib_ij_i} \mid I_{(t_i-1)b_i\cdot}) \big]^{I_{t_ib_ij_i}},$}
\]
so that the complete-data log-likelihood is
$$ l = \sum_{i=1}^{N} \sum_{t_ib_i} \sum_{j_i=1}^{n_i+1} I_{t_ib_ij_i}\Big[ \log P(D_{t_ib_i} \mid I_{t_ib_ij_i}) + \log P(\eta_{t_ib_i} \mid I_{t_ib_ij_i}) + \log P(I_{t_ib_ij_i} \mid I_{(t_i-1)b_i\cdot}) \Big], $$
where for an engaged state $j_i \le n_i$ the positional term expands to $-\tfrac{1}{2\sigma^2_D}\lVert D_{t_ib_i} - \gamma_r R_{t_ij_i} - \gamma_q Q_{t_i}\rVert^2 - \log\sigma^2_D$ up to a constant, and for the disengaged state the two emission terms are the constants $c_{\text{bg}}$ and $-\log 2\pi$. 

We group the parameters into two blocks: the emission and matchup-prior block $\Theta_1 = (\gamma_r, \gamma_q, \sigma^2_D, \beta)$, and the transition block $\Theta_2 = (\{\rho_b, p_{\text{fail},b}\}_b, \rho_{\text{null}})$. The constants $\kappa$, $c_{\text{bg}}$, and $\nu$ are fixed.

\subsection{Inference} \label{sec:inference}
The two blocks are estimated by a two-stage approach rather than a single joint EM. \textbf{Stage 1} fits $\Theta_1$ by EM on the engaged-only model, obtained by deleting the disengaged state. As such, each chain has $n_i$ states, the transition reduces to the stay/switch form $P(I_{tbj} \mid I_{(t-1)bj}) = \rho$, $P(I_{tbj} \mid I_{(t-1)bj'}) = (1-\rho)/(n_i - 1)$, and the initial distribution is the conditional-logit prior over rushers alone ($\nu = 0$). Stage 1 also produces a preliminary per-blocker stickiness under a Beta-Binomial prior. \textbf{Stage 2} then holds $\hat\Theta_1$ fixed, restores the disengaged state, and fits $\Theta_2$ by EM on the full $(n_i+1)$-state model. Each stage runs a fixed $15$ EM iterations.

We hold $\Theta_1$ fixed during estimation of $\Theta_2$ since this allows the model to be identified. Under the augmented model, the background density $c_{\text{bg}}$ and the emission distribution variance $\sigma^2_D$ can both be used for explaining poorly-fit frames. For example, a smaller variance would increase the prevalence of disengaged states, and vice versa. 

\subsubsection{Stage 1: EM on the Engaged-Only Model}
Utilizing the Expectation Maximization (EM) algorithm, we estimate the unknown $I_{t_ib_ij_i} \forall i = 1 \dots N$ (with $j_i \le n_i$), $\sigma^2_D$, $\gamma_r$, $\gamma_q$ and $\rho$. At each iteration $s$ of the algorithm, we perform the E-step and M-step. For the E-step, we compute $E_{t_ib_ij_i}^{(s)} = E[I^{(s)}_{t_ib_ij_i} | D_{t_ib_i}, \hat{\rho^{(s)}}, \hat{\sigma^{2(s)}_D}, \hat{\Gamma^{(s)}}] $ and $A_{t_ib_ij_ij_{i}^{'}} = E[I_{t_i b_i j_i} I_{(t_i-1) b_i j_i^{'} }| D_{t_ib_i}, \hat{\rho^{(s)}}, \hat{\sigma^{2(s)}_D}, \hat{\Gamma^{(s)}}]$ $\forall i = 1 \dots N$
where $\Gamma = [\gamma_r, \gamma_q]$.
Using the forward-backward algorithm, we can compute these expectations by each $b_i$ since we assume the pass blocking assignments are independent across blockers.
In the M-step, we update the maximum likelihood estimates of $\sigma^2_D$, $\Gamma$, and $\rho$ given the current expectations. 
Letting $\mathbf{X} = [\mathbf{R}, \mathbf{Q}]$ be the design matrix corresponding to the rusher location and quarterback location. Then $X_{t_ij_i} = [R_{t_ij_i}, Q_{t_i}]$ be the design matrix corresponding to rusher $j$ during play $i$ at time $t_i$.
In the $s$th iteration of the M-step we first update estimates of $\Gamma$ and $\sigma^2_D$
$$ (\hat{\Gamma^{(s)}}, \hat{\sigma^{2(s)}_D}) \xleftarrow{} \underset{\Gamma \,:\, \mathbf{1}^\top\Gamma = 1,\; \sigma^2_D}{\mathrm{argmax}} \; - \sum_{i=1}^{N} \sum_{t_i b_ij_i} E_{t_ib_ij_i}^{(s)}\Big[\frac{1}{2\sigma^2_D}\lVert D_{t_ib_i} - \Gamma X_{t_ij_i}\rVert^2 + \log(\sigma^2_D)\Big]$$
The estimator of $\Gamma$ is the responsibility-weighted generalized least-squares solution projected onto the constraint $\gamma_r + \gamma_q = 1$, as in \cite{franks2015characterizing}.
Note that the estimated pass blocker variation at time step $s$, $\hat{\sigma}^2_D$, is the weighted residual sum of squares at the updated $\hat{\Gamma}$ divided by the total responsibility mass,
$$ \hat{\sigma}^2_D = \frac{\sum_{i=1}^N \mathrm{tr}\big[(D_i - \hat{\Gamma}X_i)^{T}\mathcal{E}_i(D_i - \hat{\Gamma}X_i)\big]}{2\sum_{i=1}^N \mathrm{tr}(\mathcal{E}_i)} $$
where $\mathcal{E}_i = diag(E_{t_ib_ij_i}^{(s)})$ over the rows of $X_i$ and the residual $D_i - \hat{\Gamma}X_i$ has one column per field dimension. Because the emission is isotropic, the residuals from both dimensions are pooled into this single variance.
Next, we update our estimate of the transition parameter, $\rho$, in iteration $s$:
$$ \hat{\rho}^{(s)} \xleftarrow{} \underset{\rho}{\mathrm{argmax}} \; \Big(\sum_{i=1}^{N} \sum_{t_i b_i j_i} A_{t_ib_ij_ij_i}\Big) log(\rho) + \Big(\sum_{i=1}^{N} \sum_{t_i b_i j_i} \sum_{j_i \neq j_i^{'}} A_{t_ib_ij_ij_i^{'}}\Big) log\Big(\frac{1-\rho}{n_i - 1}\Big) $$
Since the $log(n_i - 1)$ term does not depend on $\rho$, the maximum likelihood estimate is the pooled fraction of retained assignments,
$$ \hat{\rho}^{(s)} = \frac{S}{S + W} = \frac{\hat{Q}}{1 + \hat{Q}}, \qquad \hat{Q} = \frac{S}{W}, $$
where $S = \sum_{i=1}^{N} \sum_{t_i b_i j_i} A_{t_ib_ij_ij_i}$ and $W = \sum_{i=1}^{N} \sum_{t_i b_i j_i} \sum_{j_i \neq j_i^{'}} A_{t_ib_ij_ij_i^{'}}$ are the total expected numbers of retained and switched assignments, respectively, and $n_i$ is the number of rushers in play $i$. Every frame-to-frame transition contributes equally to the estimate.

\paragraph{Transition Heterogeneity By Blocker}
The pooled estimate above assigns one parameter to an entire position group. We instead give each blocker $b$ his own $\rho_b$ with a shared conjugate prior $\rho_b \sim \text{Beta}(\alpha_\rho, \beta_\rho)$, and produce the analogous sufficient statistics in the pooled case from $A$ into per-blocker expected transition counts
$$ S_b = \sum_{i, t_i, j_i} A_{t_i b j_i j_i}, \qquad W_b = \sum_{i, t_i, j_i} \sum_{j_i \neq j_i^{'}} A_{t_i b j_i j_i^{'}}, $$
. Setting $\alpha_\rho = \rho_0 c$ and $\beta_\rho = (1-\rho_0) c$ to a fixed weakly informative prior recovers an ordinary conjugate-Bayes update in which the per-blocker step is fully closed form. Conditional on $\rho_b$, each frame-to-frame transition is a Bernoulli ``stay versus switch'' trial. The posterior is $\rho_b \mid S_b, W_b \sim \text{Beta}(\alpha_\rho + S_b,\, \beta_\rho + W_b)$, and the maximization occurs at the mean
$$ \hat{\rho}_b = \frac{S_b + \alpha_\rho}{S_b + W_b + \alpha_\rho + \beta_\rho}. $$
This allows a blocker with many transitions to be influenced primarily by his own counts $S_b, W_b$, while one with few is shrunk toward the prior mean $\alpha_\rho/(\alpha_\rho+\beta_\rho)$.

 Each EM iteration thus updates every $\rho_b$ in closed form, after which the expected counts $S_b, W_b$ are recomputed at the next E-step. Within Stage 1 the per-blocker $\rho_{b}$ replaces the shared $\rho$ when forming the transition matrix in the E-step. They are re-estimated in Stage 2 under the structured transition, for which the Stage-1 $\hat\rho_b$ and prior mean $\rho_0 = \alpha_\rho/(\alpha_\rho + \beta_\rho)$ serve as the starting point. 

\paragraph{Matchup Prior}
The conditional-logit coefficients $\beta$ of the initial distribution have no closed-form update. In the M-step we maximize the corresponding term of the expected complete-data log-likelihood,
$$ Q(\beta) = \sum_{i=1}^{N} \sum_{b_i} \sum_{j_i} E_{0 b_i j_i}^{(s)} \; log\Big( \frac{\exp\{\beta^\top X_{0 b_i j_i}\}}{\sum_{j_i^{'}} \exp\{\beta^\top X_{0 b_i j_i^{'}}\}} \Big), $$
where $E_{0 b_i j_i}^{(s)}$ are the snap-frame ($t=0$) responsibilities used as soft targets. This is a soft-label multinomial-logistic regression with gradient
$$ \nabla_\beta Q = \sum_{i=1}^{N} \sum_{b_i} \sum_{j_i} \big( E_{0 b_i j_i}^{(s)} - P(I_{0 b_i j_i} = 1) \big) X_{0 b_i j_i}; $$
we maximize it with a small number of Newton steps warm-started from the previous iterate. The pre-snap features $X_{0 b_i j_i}$ are standardized and $\beta$ is shared across plays but estimated separately per blocking position.

\subsubsection{Stage 2: Disengaged Transition Estimation}
Stage 2 restores the $(n_i+1)$-th disengaged state and its structured transition (Model Formulation above) and estimates $\Theta_2 = (\{\rho_b, p_{\text{fail},b}\}_b, \rho_{\text{null}})$, holding the Stage-1 emission weights, variance, matchup-prior coefficients, and feature standardization fixed. The transition is initialized at the Stage-1 per-blocker $\hat\rho_b$, with $p_{\text{fail},b} = 0.01$ and $\rho_{\text{null}} = 0.99$.

Each iteration runs the forward--backward algorithm over the $n_i + 1$ states, with the engaged emissions evaluated at $\hat\Theta_1$, the disengaged emission at $c_{\text{bg}}$, and the initial distribution with fixed disengaged probability $\nu$. The posteriors yield expected counts per blocker, with the following sufficient statistics: $S_b$ (stay on the same rusher), $W_b$ (switch to another rusher), $F_b$ (engaged$\to$disengaged), and disengaged-row counts $N^{\text{stay}}$ (disengaged$\to$disengaged) and $N^{\text{re}}$ (disengaged$\to$rusher) pooled over the position group. The sufficient statistics $(S_b, W_b, F_b)$ are multinomial, so we place a shared Dirichlet$(\alpha_1, \alpha_2, \alpha_3)$ prior over their three outcomes  as $\alpha = (20\rho_0,\, 18(1-\rho_0),\, 2(1-\rho_0))$. The per-blocker M-step is then closed form,
$$ \hat{\rho}_b = \frac{S_b + \alpha_1}{S_b + W_b + F_b + \alpha_0}, \qquad \hat{p}_{\text{fail},b} = \frac{F_b + \alpha_3}{S_b + W_b + F_b + \alpha_0}, \qquad \alpha_0 = \alpha_1 + \alpha_2 + \alpha_3, $$
and the disengaged persistence is the pooled fraction, $\hat{\rho}_{\text{null}} = (N^{\text{stay}} + 1) / (N^{\text{stay}} + N^{\text{re}} + 2)$. These Stage-2 estimates form the transition matrix used to produce every assignment posterior, both the smoothed forward--backward posterior and the filtered posterior of the continuous-time model.

\FloatBarrier
\section{Play-Level Plus-Minus Model} \label{sec:playlevel}
The opponent-quality covariate $R_j$ in the block-hold hazard of Section~\ref{shedding} is the rusher effect of a play-level plus-minus model, which the continuous-time model of Section~\ref{continuous} extends to the frame level. The model quantifies the maximum pressure a rusher generates on a play, while leveraging the partial responsibilities of the pass-blockers generated from the HMM. For each rusher $j$ on play $i$, we take $y_{ij} = \max_t \text{STRAIN}_{ijt}$, the maximum STRAIN that rusher exerts over the frames of the play. We model a rusher's peak STRAIN as that rusher's effect $R_j$, subtracted by the contribution of the blockers $B_b$ assigned to him, weighted by the average attention $\theta(b,j,i)$ that blocker $b$ devotes to rusher $j$ on play $i$. Because of this formulation we capture the partial contributions of all blockers on a play, and a more engaged blocker (larger $B_b$) reduces the STRAIN its rusher can generate.
$$ y_{ij} \sim \mathcal{N}\Big( \mu + R_{j} - \!\!\sum_{b \in \mathcal{X}_i}\!\! B_{b}\,\theta(b, j, i) + Q_{q_i} + O_{o_i} + D_{d_i} + \mathrm{Dn}_{i} + \mathrm{Qt}_{i} + \mathrm{OF}_{i} + \mathrm{DC}_{i} + \gamma^\top s_i , \; \sigma_y^2 \Big) $$
where $\mathrm{Dn}_i, \mathrm{Qt}_i, \mathrm{OF}_i, \mathrm{DC}_i$ are random intercepts for the down, quarter, offensive formation, and defensive coverage type of play $i$, $Q_{q_i}, O_{o_i}, D_{d_i}$ are random intercepts for the quarterback, offensive team, and defensive team, and $s_i$ denotes covariates of score differential, game time remaining, and field position.

The rusher, blocker, and quarterback effects are drawn around a linear function of player attributes $z = [\text{position}, \text{height}, \text{weight}]$, which shrinks each player toward their positional/size archetype, while the team and game-situation intercepts are zero-centered:
$$ R_j = \alpha_R^\top z_j + \sigma_R\,\varepsilon^R_j, \quad B_b = \alpha_B^\top z_b + \sigma_B\,\varepsilon^B_b, \quad Q_q = \alpha_Q^\top z_q + \sigma_Q\,\varepsilon^Q_q, $$
$$ O_o = \sigma_{\mathrm{Off}}\,\varepsilon^O_o,\quad D_d = \sigma_{\mathrm{Def}}\,\varepsilon^D_d,\quad \mathrm{Dn}_i = \sigma_{\mathrm{Dn}}\,\varepsilon^{\mathrm{Dn}}_{\mathrm{dn}(i)}, \quad\text{and likewise } \mathrm{Qt},\,\mathrm{OF},\,\mathrm{DC}, \qquad \varepsilon \sim \mathcal{N}(0,1). $$
The intercept $\mu$, the attribute-coefficient vectors $\alpha_R, \alpha_B, \alpha_Q$, and the game context effects $\gamma$ are given $\mathcal{N}(0,1)$ priors, and every scale parameter --- the residual $\sigma_y$ together with $\sigma_R, \sigma_B, \sigma_Q, \sigma_{\mathrm{Off}}, \sigma_{\mathrm{Def}}$ and the four game context scales $\sigma_{\mathrm{Dn}}, \sigma_{\mathrm{Qt}}, \sigma_{\mathrm{OF}}, \sigma_{\mathrm{DC}}$ --- is given an $\text{Inverse-Gamma}(2,1)$ prior. All continuous covariates are standardized.

In contrast to previous plus/minus metrics which incorporate every on-field participant, we include only the rusher of interest, excluding the other rushers. We argue instead that the blocking assignment $\theta(\cdot, j, i)$ is a sufficient conditioning variable for the influence of the remaining rushers. The only way another rusher $k$ can affect $j$'s STRAIN is by competing for the finite blocking resource and drawing assignment mass away from $j$. Recall that for each blocker $\sum_{j} \theta(b,j,i) = 1$ and that the reallocation is fully observed in the $\theta(b,j,i)$ that $j$ receives. Conditional on $\theta(\cdot, j, i)$, the identities and abilities of the other rushers therefore carry no additional information about $j$'s STRAIN.

We fit the play-level model on all eight weeks of data ($8{,}530$ plays, $36{,}254$ rusher observations) with the No-U-Turn sampler \citep{hoffman2014no} implemented using NumPyro \citep{phan2019composable}, running four chains of $1{,}000$ warmup and $2{,}000$ retained iterations each ($8{,}000$ posterior draws). Convergence diagnostics are reported in Appendix~\ref{sec:diagnostics}.

\FloatBarrier
\section{Posterior Sampling Diagnostics} \label{sec:diagnostics}
Three of the models in this paper are fit by the No-U-Turn sampler: the play-level plus-minus model of Appendix~\ref{sec:playlevel}, the continuous-time blocker rating of Section~\ref{continuous}, and the opponent-adjusted block-hold hazard of Section~\ref{shedding}. Each was run as four chains from dispersed initializations --- $1{,}000$ warmup and $2{,}000$ retained iterations per chain for the play-level model, $800$ and $800$ for the continuous-time model, and $1{,}000$ and $1{,}000$ for the block-hold model.
Table~\ref{tab:mcmc_diagnostics} reports $\widehat{R}$ and the bulk and tail effective sample sizes for the parameters of interest.
\begin{table}[h!]
\centering
\footnotesize
\begin{tabular}{lcccc}
\toprule
Parameter & Components & $\widehat{R}$ & ESS (bulk) & ESS (tail) \\
\midrule
\multicolumn{5}{l}{\emph{Play-level plus-minus} (4 chains)} \\
\midrule
$\mu$ (intercept) & 1 & 1.002 & 3,107 & 3,827 \\
$R_j$ (rusher effects) & 698 & 1.002 & 4,076 & 3,952 \\
$B_b$ (blocker effects) & 534 & 1.003 & 9,853 & 4,796 \\
$Q_q$ (quarterback effects) & 60 & 1.002 & 6,683 & 4,552 \\
$\alpha_R,\alpha_B,\alpha_Q$ (attributes) & 22 & 1.002 & 2,583 & 3,908 \\
$\sigma_R,\sigma_B,\sigma_Q$ & 3 & 1.002 & 2,816 & 4,470 \\
$\sigma_{\mathrm{Off}},\sigma_{\mathrm{Def}}$ & 2 & 1.001 & 3,848 & 5,263 \\
$\sigma_{\mathrm{Dn}},\sigma_{\mathrm{Qt}},\sigma_{\mathrm{OF}},\sigma_{\mathrm{DC}}$ & 4 & 1.001 & 3,549 & 4,539 \\
$\gamma$ (situation slopes) & 3 & 1.001 & 11,456 & 4,851 \\
$\sigma_y$ (residual) & 1 & 1.000 & 12,877 & 4,944 \\
\midrule
\multicolumn{5}{l}{\emph{Continuous-time} (4 chains)} \\
\midrule
$\mu$ (intercept) & 1 & 1.000 & 2,311 & 2,574 \\
$B^{\Delta}_b$ (blocker effects) & 534 & 1.007 & 2,333 & 1,601 \\
$\alpha_B$ (blocker attributes) & 8 & 1.001 & 2,331 & 2,336 \\
$\sigma_B^{\Delta}$ & 1 & 1.003 & 1,318 & 2,021 \\
$R^{\Delta}_j$ (rusher effects) & 698 & 1.006 & 1,610 & 1,591 \\
$\alpha_R$ (rusher attributes) & 13 & 1.002 & 3,055 & 2,236 \\
$\sigma_R^{\Delta}$ & 1 & 1.003 & 1,054 & 1,266 \\
$Q^{\Delta}_q$ (QB effects) & 60 & 1.005 & 2,219 & 1,888 \\
$\alpha_Q$ (QB attributes) & 2 & 1.003 & 1,700 & 1,866 \\
$\sigma_Q^{\Delta}$ & 1 & 1.000 & 1,224 & 1,734 \\
$\phi$ (STRAIN$_t$ control) & 1 & 1.002 & 3,120 & 2,158 \\
$\sigma_{\Delta}$ (residual) & 1 & 1.003 & 3,470 & 2,110 \\
\midrule
\multicolumn{5}{l}{\emph{Block hold} (4 chains)} \\
\midrule
$\alpha$ (baseline hazard) & 1 & 1.004 & 944 & 1,579 \\
$g_1,g_2$ (dwell) & 2 & 1.002 & 5,664 & 3,074 \\
$\beta_{\mathrm{opp}}$ (opponent quality) & 1 & 1.000 & 6,719 & 2,443 \\
$\sigma_u$ (frailty scale) & 1 & 1.001 & 978 & 1,926 \\
$-u_b$ (hold ratings) & 534 & 1.006 & 3,220 & 1,871 \\
\bottomrule
\end{tabular}
\caption{Convergence diagnostics for the three models fit by the No-U-Turn sampler, each run as four chains. For vector parameters, Components gives the number of scalar components and the table reports the \emph{worst} value over them (largest $\widehat{R}$, smallest effective sample size). All parameters satisfy $\widehat{R} \leq 1.007$, with bulk and tail effective sample sizes of at least 944.}
\label{tab:mcmc_diagnostics}
\end{table}

\clearpage
\section{Supplementary Tables}
These supplementary tables illustrate the variety of metric rankings derived from our models.
\subsection{Attention Metrics}

\begin{table}[h!]
\centering
\begin{subtable}{0.24\textwidth}
\centering
\scriptsize
\begin{tabular}{lc}
\toprule
Name & Att. \\
\midrule
K'Lavon Chaisson & 0.72 \\
Justin Hollins & 0.75 \\
Von Miller & 0.77 \\
Dont'a Hightower & 0.79 \\
Terrell Lewis & 0.80 \\
\bottomrule
\end{tabular}
\caption{Edge}
\end{subtable}
\hfill
\begin{subtable}{0.24\textwidth}
\centering
\scriptsize
\begin{tabular}{lc}
\toprule
Name & Att. \\
\midrule
Chris Jones & 1.22 \\
Zach Sieler & 1.23 \\
Dalvin Tomlinson & 1.23 \\
Michael Hoecht & 1.23 \\
Ross Blacklock & 1.24 \\
\bottomrule
\end{tabular}
\caption{DT}
\end{subtable}
\hfill
\begin{subtable}{0.24\textwidth}
\centering
\scriptsize
\begin{tabular}{lc}
\toprule
Name & Att. \\
\midrule
Josh Tupou & 1.29 \\
Poona Ford & 1.31 \\
Danny Shelton & 1.31 \\
Steve McLendon & 1.35 \\
Eddie Goldman & 1.35 \\
\bottomrule
\end{tabular}
\caption{NT}
\end{subtable}
\caption{Bottom 5 pass rushers by front-normalized attention, by position (min 50 snaps).}
\label{tab:att_bot_norm}
\end{table}

\subsection{Disengagement Metrics}

\begin{figure}[h!]
\centering
\includegraphics[width=\textwidth]{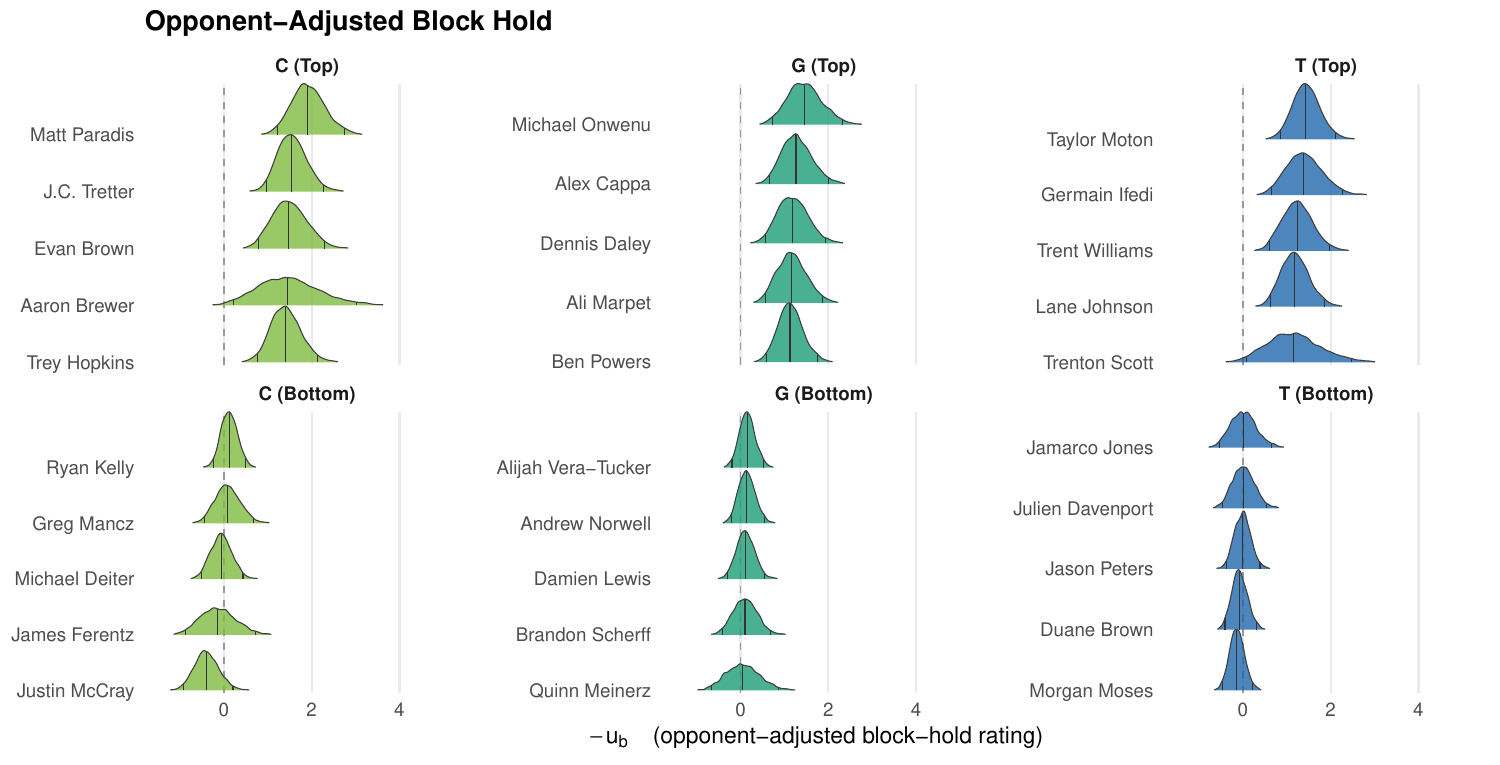}
\caption{Posterior distributions of the opponent-adjusted block-hold rating $-u_b$ for the top and bottom five pass blockers by position (min 40 blocks). Higher values indicate a blocker who sustains his block longer.}
\label{fig:block_hold}
\end{figure}

\begin{table}[h!]
\centering
\begin{subtable}{0.48\textwidth}
\centering
\scriptsize
\begin{tabular}{lccc}
\toprule
Name & RMST (s) & Beat rate & Adj resid (s) \\
\midrule
Joe Haeg & 4.000 & 0.014 & -1.451 \\
Germain Ifedi & 3.966 & 0.031 & -0.708 \\
Lane Johnson & 3.965 & 0.061 & -0.340 \\
Cornelius Lucas & 3.961 & 0.054 & -0.616 \\
Braden Smith & 3.961 & 0.068 & -0.706 \\
Lucas Niang & 3.956 & 0.061 & -0.583 \\
Rashawn Slater & 3.956 & 0.040 & -0.983 \\
Trenton Scott & 3.956 & 0.023 & -0.798 \\
Dan Moore & 3.954 & 0.030 & -1.279 \\
Marcus Cannon & 3.953 & 0.057 & -0.829 \\
Olisaemeka Udoh & 3.946 & 0.035 & -0.968 \\
Andrew Thomas & 3.945 & 0.030 & -0.947 \\
Taylor Moton & 3.944 & 0.031 & -0.748 \\
Matt Nelson & 3.943 & 0.049 & -0.885 \\
Trent Williams & 3.942 & 0.040 & -0.874 \\
\bottomrule
\end{tabular}
\caption{Tackle pass-block engagement (longer survival-to-beat = better)}
\label{tab:block_engagement}
\end{subtable}
\hfill
\begin{subtable}{0.48\textwidth}
\centering
\scriptsize
\begin{tabular}{lccc}
\toprule
Name & Pos & RMST (s) & Beat rate \\
\midrule
Dont'a Hightower & Edge & 3.663 & 0.161 \\
Jonathan Greenard & Edge & 3.671 & 0.140 \\
Michael Hoecht & DT & 3.674 & 0.150 \\
Alton Robinson & Edge & 3.682 & 0.172 \\
Yetur Gross-Matos & Edge & 3.685 & 0.143 \\
Zach Sieler & DT & 3.706 & 0.122 \\
Chauncey Golston & Edge & 3.721 & 0.155 \\
Andrew Van Ginkel & Edge & 3.733 & 0.113 \\
Nick Bosa & Edge & 3.737 & 0.113 \\
Lawrence Guy & DT & 3.744 & 0.141 \\
A.J. Epenesa & Edge & 3.748 & 0.160 \\
Danny Shelton & NT & 3.749 & 0.135 \\
Steven Means & Edge & 3.750 & 0.147 \\
Carl Davis & NT & 3.754 & 0.095 \\
Chase Winovich & Edge & 3.756 & 0.122 \\
\bottomrule
\end{tabular}
\caption{Pass rushers that beat blocks fastest (shortest survival-to-beat)}
\label{tab:rusher_shedding}
\end{subtable}
\caption{Two sides of the same block-failure survival outcome: tackles ranked by how long they sustain a block (a) and pass rushers by how quickly they beat one (b); min 40 blocks.}
\label{tab:beat_rates}
\end{table}

\subsection{Plus Minus Metrics}

\begin{figure}[h!]
\centering
\includegraphics[width=\textwidth]{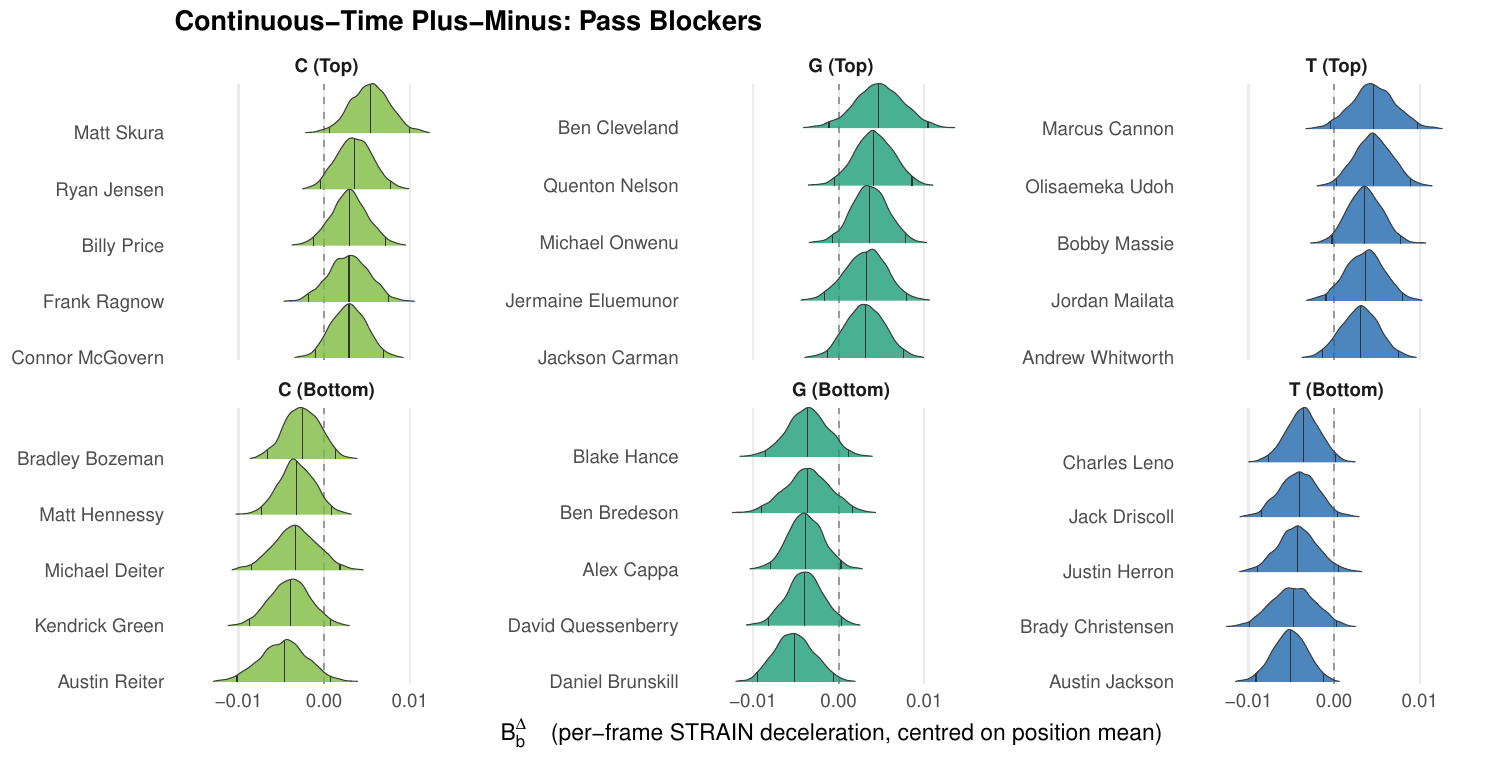}
\caption{Posterior distributions of the continuous-time blocker effect $B^{\Delta}_b$ for the top and bottom five pass blockers by position (min 50 pass-block snaps). Higher values indicate a blocker whose engagement decelerates the rush more. Zero is the average player at the position.}
\label{fig:blocker_pm}
\end{figure}

\subsection{Quarterback STRAIN Metrics}
\begin{figure}[h!]
\centering
\includegraphics[width=0.72\textwidth]{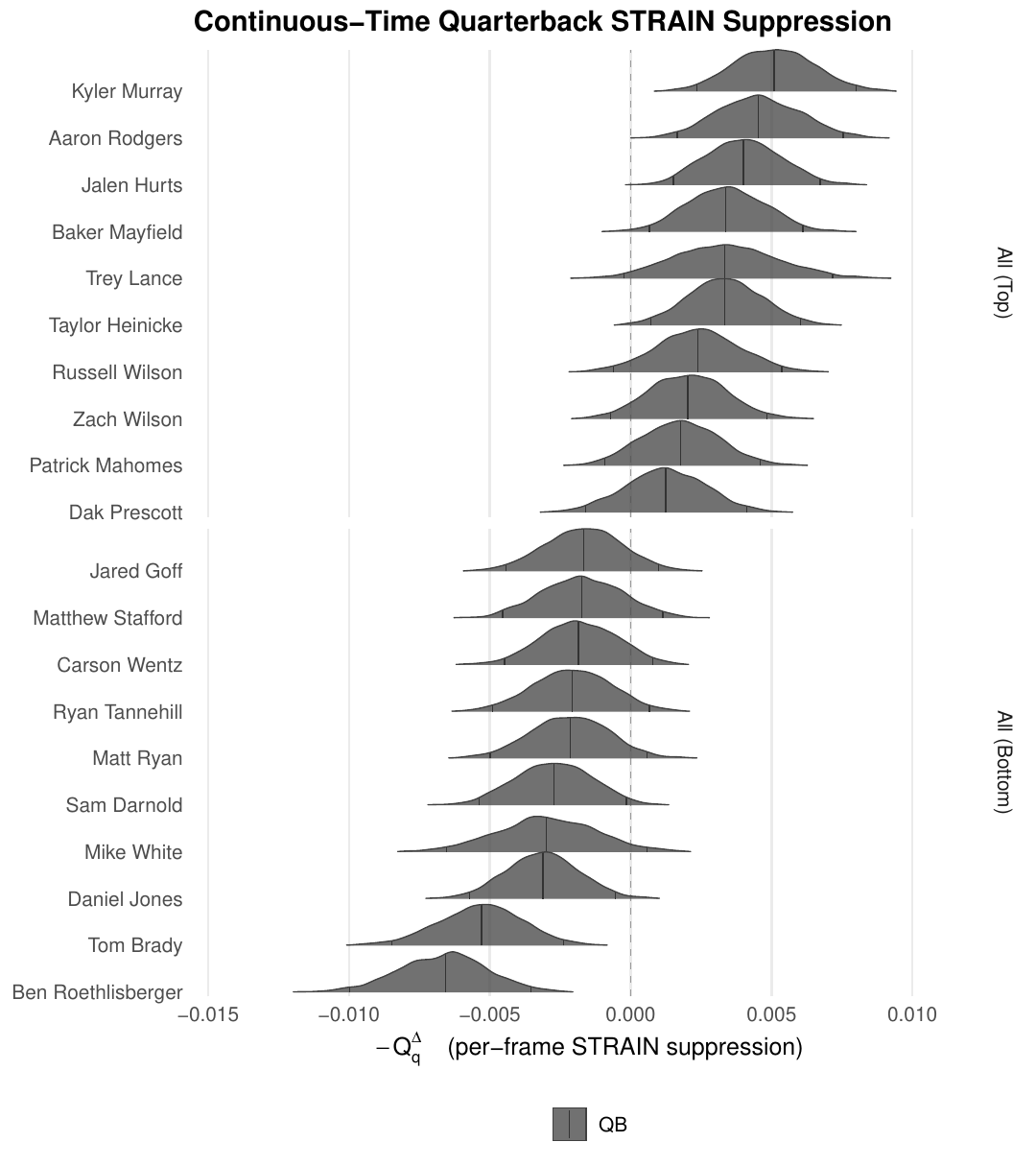}
\caption{Posterior distributions of continuous-time quarterback STRAIN suppression $-Q^{\Delta}_q$ for the ten most and ten least suppressing passers (min 50 dropbacks). Higher values indicate slower per-frame STRAIN growth on that quarterback's dropbacks.}
\label{fig:qb_suppression}
\end{figure}

Because the quarterback intercept $Q^{\Delta}_{q_i}$ enters the continuous-time predictor of Section~\ref{continuous} additively, the model yields a by-product ranking of quarterbacks by how quickly STRAIN builds on their dropbacks. Reporting this as a suppression score $-Q^{\Delta}_q$, Figure~\ref{fig:qb_suppression} lists the most and least STRAIN-suppressing passers. Those who suppress the most (Kyler Murray, Aaron Rodgers, and Jalen Hurts) are known for their mobility, whereas STRAIN builds fastest on the dropbacks of the less mobile Ben Roethlisberger and Tom Brady. We note that this score reflects STRAIN rather than sacks, so it measures how quickly pressure develops rather than whether it is converted. The posterior distributions are wide and overlap, so only the extremes of the ranking are clearly separated.

\subsection{Pocket Space Control Metrics}
The pocket space-control tables of Section~\ref{pocketspace} follow: the per-rusher space-won ranking and the assignment-weighted space-generation-gain table.

\begin{table}[h!]
\centering
\begin{subtable}{0.48\textwidth}
\centering
\footnotesize
\begin{tabular}{lcc}
\toprule
Name & Space won & 95\% CI \\
\midrule
Preston Smith & 12.30 & $[11.80,\,12.89]$ \\
Isaac Rochell & 12.17 & $[10.64,\,14.07]$ \\
Darrell Taylor & 12.17 & $[11.57,\,12.93]$ \\
Genard Avery & 12.16 & $[11.00,\,13.42]$ \\
John Franklin-Myers & 12.13 & $[11.59,\,12.77]$ \\
\bottomrule
\end{tabular}
\caption{Edge}
\end{subtable}
\hfill
\begin{subtable}{0.48\textwidth}
\centering
\footnotesize
\begin{tabular}{lcc}
\toprule
Name & Space won & 95\% CI \\
\midrule
Broderick Washington & 9.61 & $[8.73,\,10.68]$ \\
Michael Hoecht & 9.33 & $[8.80,\,9.80]$ \\
AShawn Robinson & 9.14 & $[8.69,\,9.63]$ \\
Chris Jones & 9.10 & $[8.65,\,9.57]$ \\
Leonard Williams & 9.07 & $[8.71,\,9.33]$ \\
\bottomrule
\end{tabular}
\caption{DT}
\end{subtable}

\bigskip
\begin{subtable}{0.48\textwidth}
\centering
\footnotesize
\begin{tabular}{lcc}
\toprule
Name & Space won & 95\% CI \\
\midrule
Josh Tupou & 8.80 & $[8.31,\,9.24]$ \\
Raekwon Davis & 8.69 & $[7.75,\,9.65]$ \\
Eddie Goldman & 8.45 & $[7.94,\,8.94]$ \\
Sebastian Joseph & 8.41 & $[8.11,\,8.70]$ \\
Damion Square & 8.41 & $[7.93,\,8.88]$ \\
\bottomrule
\end{tabular}
\caption{NT}
\end{subtable}
\hfill
\begin{subtable}{0.48\textwidth}
\centering
\footnotesize
\begin{tabular}{lcc}
\toprule
Name & Space won & 95\% CI \\
\midrule
David Long & 11.80 & $[10.82,\,12.82]$ \\
Elandon Roberts & 11.57 & $[10.13,\,13.11]$ \\
Roquan Smith & 11.48 & $[10.25,\,12.64]$ \\
Kenny Young & 11.42 & $[10.68,\,12.19]$ \\
Deion Jones & 11.35 & $[10.29,\,12.48]$ \\
\bottomrule
\end{tabular}
\caption{ILB}
\end{subtable}

\bigskip
\end{table}

\begin{table}[h!]
\centering
\begin{subtable}{0.48\textwidth}
\centering
\footnotesize
\begin{tabular}{lcc}
\toprule
Name & Space won & 95\% CI \\
\midrule
Jamin Davis & 11.47 & $[10.14,\,12.86]$ \\
Azeez Al-Shaair & 10.48 & $[8.78,\,12.33]$ \\
Tremaine Edmunds & 10.13 & $[8.97,\,11.44]$ \\
Myles Jack & 10.13 & $[8.80,\,11.40]$ \\
Shaquille Leonard & 9.93 & $[8.96,\,10.97]$ \\
\bottomrule
\end{tabular}
\caption{MLB}
\end{subtable}
\hfill
\begin{subtable}{0.48\textwidth}
\centering
\footnotesize
\begin{tabular}{lcc}
\toprule
Name & Space won & 95\% CI \\
\midrule
Jabrill Peppers & 11.59 & $[10.02,\,13.58]$ \\
Jamal Adams & 11.34 & $[10.21,\,12.79]$ \\
Malcolm Jenkins & 9.97 & $[9.01,\,10.89]$ \\
Chuck Clark & 8.64 & $[7.66,\,9.47]$ \\
Brandon Jones & 8.61 & $[7.87,\,9.42]$ \\
\bottomrule
\end{tabular}
\caption{SS}
\end{subtable}

\bigskip
\begin{subtable}{0.48\textwidth}
\centering
\footnotesize
\begin{tabular}{lcc}
\toprule
Name & Space won & 95\% CI \\
\midrule
Jeremy Chinn & 10.98 & $[9.37,\,12.76]$ \\
Jayron Kearse & 10.79 & $[9.16,\,12.59]$ \\
Daniel Sorensen & 9.16 & $[8.12,\,10.24]$ \\
\bottomrule
\end{tabular}
\caption{FS}
\end{subtable}
\hfill
\begin{subtable}{0.48\textwidth}
\centering
\footnotesize
\begin{tabular}{lcc}
\toprule
Name & Space won & 95\% CI \\
\midrule
Mike Hilton & 12.76 & $[11.06,\,14.54]$ \\
LJarius Sneed & 11.63 & $[10.69,\,12.72]$ \\
\bottomrule
\end{tabular}
\caption{CB}
\end{subtable}

\bigskip
\caption{Per-rusher pocket space-won (control share over the QB danger region, yd$^2$; higher = collapses more of the QB's space), by position, bootstrap 95\% CI over plays.}
\label{tab:pocket_rusher_space}
\end{table}

\begin{table}[h!]
\centering
\footnotesize
\begin{tabular}{llcccc}
\toprule
Name & Pos & SGG & Att. & PFF press & $+/-$ \\
\midrule
Grover Stewart & DT & 1.129 & 1.30 & 0.07 & -0.17 \\
Danny Shelton & NT & 1.117 & 1.23 & 0.08 & -0.17 \\
Alim McNeill & NT & 0.913 & 1.25 & 0.05 & -0.19 \\
Austin Johnson & NT & 0.896 & 1.22 & 0.10 & -0.12 \\
Bryan Mone & NT & 0.854 & 1.28 & 0.04 & -0.24 \\
Robert Nkemdiche & DT & 0.841 & 1.26 & 0.05 & -0.12 \\
Sheldon Rankins & DT & 0.830 & 1.31 & 0.08 & -0.11 \\
Kingsley Keke & Edge & 0.750 & 1.24 & 0.06 & -0.19 \\
Justin Hollins & Edge & 0.741 & 0.62 & 0.04 & -0.13 \\
Dalvin Tomlinson & DT & 0.723 & 1.11 & 0.12 & -0.14 \\
Romeo Okwara & Edge & 0.713 & 0.76 & 0.21 & -0.07 \\
Teair Tart & DT & 0.709 & 1.36 & 0.05 & -0.15 \\
Dorance Armstrong & Edge & 0.705 & 0.96 & 0.12 & -0.09 \\
Wyatt Ray & Edge & 0.703 & 0.88 & 0.10 & -0.15 \\
Davon Godchaux & NT & 0.700 & 1.26 & 0.04 & -0.15 \\
\bottomrule
\end{tabular}
\caption{Top pass rushers by Space Generation Gain (SGG). Att.\ is the rusher's front-normalized blocking attention, PFF press his own pressure rate, and $+/-$ his plus-minus effect. The leading generators are interior linemen.}
\label{tab:pocket_sgg}
\end{table}

\FloatBarrier
\end{appendices}
\clearpage

\begin{acknowledgement}
  Big Data Bowl for publicly-available data
\end{acknowledgement}

\bibliographystyle{chicago}
\bibliography{refs}

\end{document}